\documentclass[preprintnumbers,twocolumn,groupedaddress,nofootinbib,aps,prb]{revtex4-1}
\usepackage{amsmath}
\usepackage{epsfig,color}
\usepackage{graphicx}
\usepackage{dcolumn}
\usepackage{bm}
\usepackage{multirow}
\usepackage{bm}
\usepackage{enumerate}
\usepackage{graphicx}
\usepackage{verbatim}
\usepackage{hyperref}
\hypersetup{colorlinks=true,linkcolor=blue,anchorcolor=blue,citecolor=blue,filecolor=blue,urlcolor=blue,bookmarksnumbered=true,pdfview=FitB}
\usepackage[utf8]{inputenc}
\usepackage[english]{babel}
\newcommand{\ve}{\varepsilon}

\newcommand{\bk}{{\bf k}}
\newcommand{\bp}{{\bf p}}
\newcommand{\bq}{{\bf q}}
\newcommand{\bQ}{{\bf Q}}

\newcommand{\nn}{\nonumber}
\newcommand{\beq}{\begin{equation}}
\newcommand{\eeq}{\end{equation}}
\newcommand{\bea}{\begin{eqnarray}}
\newcommand{\eea}{\end{eqnarray}}
\newcommand{\bse}{\begin{subequations}}
\newcommand{\ese}{\end{subequations}}
\newcommand{\bwt}{\begin{widetext}}
\newcommand{\ewt}{\end{widetext}}

\newcommand{\bv}{{\bf v}}

\newcommand{\R}{\mathrm{Re}}
\newcommand{\bsu}{\begin{subequations}}
\newcommand{\esu}{\end{subequations}}

\newcommand{\ofl}{\omega_{\text{FL}}}

\begin{document}

\title{
Optical conductivity of a two-dimensional metal near a quantum-critical point: \\
the status of the ``extended Drude formula''}
\author{Andrey V. Chubukov$^{a}$ and  Dmitrii L. Maslov$^{b}$}
\date{\today}
\affiliation{
$^{a)}$Department of Physics, University of Minnesota, Minneapolis, Minnesota 55455\\
$^{b)}$Department of Physics, University of
Florida, Gainesville, Florida 32611}
\date{\today}
\begin{abstract}
The optical conductivity of a metal near a quantum critical point (QCP)
is expected to depend on frequency not only via the scattering time but also via the
 effective mass, which acquires a singular frequency dependence near a QCP.
We check this assertion by computing
diagrammatically
the
optical conductivity, $\sigma' (\Omega)$,
 near both nematic and spin-density wave (SDW)
 quantum critical points (QCPs) in 2D.
  If renormalization of current vertices is not taken into account, $\sigma' (\Omega)$
   is expressed via the quasiparticle residue $Z$ (equal to the ratio of bare and renormalized masses in our approximation)  and
    transport scattering rate $\gamma_{\text{tr}}$ as $\sigma' (\Omega)
   \propto Z^2  \gamma_{\text{tr}}/\Omega^2$.
    For a nematic QCP ($\gamma_{\text{tr}}\propto\Omega^{4/3}$ and $Z\propto\Omega^{1/3}$), this formula suggests that $\sigma'(\Omega)$
      would tend to a constant at $\Omega  \to 0$.
        We explicitly demonstrate that
        the actual behavior of $\sigma' (\Omega)$ is different due to strong renormalization of the current vertices,
        which
     cancels out a factor of $Z^2$.
       As a result,
       $\sigma' (\Omega)$ diverges  as $
       1/\Omega^{2/3}$, as earlier works conjectured.
      In the SDW case, we consider two contributions to the conductivity: from hot spots and  from``lukewarm'' regions of the  Fermi surface.
      The hot-spot contribution is not affected by vertex renormalization, but it is subleading to the lukewarm one.  
   For the latter,  we argue that  
        a factor of $Z^2$ is again cancelled by vertex corrections. As a result,
         $ \sigma' (\Omega)$ at a SDW QCP scales as $
          1/\Omega$ down to the lowest frequencies.
 \end{abstract}

\maketitle

\section{Introduction}
\label{sec:intro}
Understanding the behavior
of fermions near a quantum-critical point (QCP) remains one of the most challenging problems
 in the physics of strongly correlated
 materials.  In dimensions $D=3$ and below, scattering by gapless excitations of
 the
  order-parameter field destroys fermionic coherence either near particular hot spots, if critical fluctuations are soft at a finite  momentum $q$, or around the entire Fermi surface (FS), if fluctuations are soft at $q=0$.  An example of a finite-$q$ QCP is a transition into a
 spin-density-wave (SDW) state, while an example of a $q=0$ QCP is a Pomeranchuk-type transition into a nematic state.  In both cases, the frequency derivative of the fermionic self-energy, $\partial \Sigma (\bk, \omega)/\partial \omega$ is
 large and singular
 near a QCP,
  and the real and imaginary parts of $\Sigma (\bk, \omega)$
  are
  of the same order.
  This violates the Landau criterion of a Fermi liquid (FL) and gives rise to a non-Fermi liquid (NFL) behavior.

Because critical behavior generally emerges
 at intermediate coupling,
there is no obvious small parameter to control a perturbation theory.
Furthermore, because soft order-parameter fluctuations are collective excitations of fermions,
the
 fermionic self-energy has to be computed self-consistently with the bosonic
 one
 (the Landau damping term)
 as both originate from the  same interactions between fermions and their collective modes.
   In $D=2$,  considered in this work,  the one-loop fermionic self-energy  due to scattering by
     critical bosons  depends predominantly on the frequency
     rather than on the momentum
      and is given by
      $\Sigma (\omega)   \propto \omega^{2/3}$  at a nematic QCP and $\Sigma (\omega) \propto \omega^{1/2}$ at a SDW QCP. In the latter case, this form holds near the hot spots (points on the FS separated by the nesting vector), in regions whose width  by itself scales as $\sqrt{\omega}$.  Higher-order terms in the loop expansion give rise to additional logarithms near both types of QCP.~\cite{abanov:2003,abanov:2004,metlitski:2010b, *metlitski:2010c,metzner:2015}
 How these logarithms modify the self-energy is not fully understood yet. We will not dwell on this issue here and use  the one-loop forms of
 the self-energy in what follows.

 The analysis of optical conductivity near a QCP brings in another level of complications.
 First,  the conductivity
 contains
a  transport
 scattering
 time which, in general, differs from the single-particle scattering
 time (given by $1/2\Sigma''$) due to constraints imposed by momentum conservation. Second,  the frequency scaling of the conductivity
 may be affected
by
 the frequency dependence of the effective mass
  near a QCP.  Phenomenologically, these two effects are often described by the \lq\lq extended
Drude formula\rq\rq,
    which has been
    widely
    used
    to analyze the
    optical
     data on the normal state of high-$T_c$ cuprates and other strongly correlated systems.
\cite{basov:2011} The most commonly used version of this formula is
    \beq
  \sigma'(\Omega)
 = \frac{\Omega^2_p}{4\pi}  \frac{\gamma_{\mathrm{tr}} (\Omega)}{\left(\Omega \frac{m^*
 }{m_b} \right)^2 + \gamma^2_{\mathrm{tr}}(\Omega)},
    \label{ch_1}
    \eeq
    where $\sigma'(\Omega)=\R\sigma(\Omega)$, $\Omega_p$ is the effective plasma frequency, $\gamma_{\mathrm{tr}}(\Omega)$ is the transport
    scattering
    rate,
    $m^*$
     is the renormalized
   effective mass which may depend on $\Omega$, and $m_b$ is the
    band mass.  This formula is motivated by the memory-matrix formalism \cite{goetze:1972} and can be viewed as a generalization
    of the usual Drude formula
     to the regime where mass renormalization is strong.
     That
     $\Omega$ is renormalized by $m^*/m_b$ can be traced down to the fact that,
      in local theories,  $m^*/m_b$ is inversely proportional to
      the
      quasiparticle
      residue
      $Z$:
    $m^*/m_b = Z^{-1} = 1 + \partial \Sigma'/\partial \omega$.
    In
 the
     cases considered in this paper, $\gamma_{\text{tr}} (\Omega)$ is smaller 
     than or at most comparable to 
     $\Omega$ at low frequencies
     (even if $\Sigma''$ is larger than $\Omega$).
      In this regime,
      Eq.~(\ref{ch_1})
      can be approximated by
     \beq
  \sigma'(\Omega)
  = \frac{\Omega^2_p}{4\pi} Z^2 \frac{\gamma_{\mathrm{tr}} (\Omega)}{\Omega^2}.
    \label{ch_1_1}
    \eeq

    In this paper, we analyze the validity of the extended Drude formula
    for two types of QCP:  a nematic one and a SDW one, both in 2D.
   We argue that,
   in general,
   this
    formula
    is incomplete and has to be modified by including
     renormalization of the current vertices,
     which is not captured by a simple replacing  of the single-particle scattering time by the transport one.

     We
     show that
     near a 2D
 nematic     QCP
     renormalization
     of the current vertices is singular  and its inclusion changes the frequency scaling of
     the
     optical conductivity,
     compared to that predicted by Eq.~(\ref{ch_1}).
That  the extended Drude formula is problematic near a 2D
nematic QCP can be readily seen by comparing the conductivity
 predicted by
Eq.~(\ref{ch_1}) with
the
result obtained
by  a
two-loop
perturbation theory in fermion-boson coupling
 \cite{kim:1994}  and
by dimensional regularization.~\cite{eberlein:2016}
As we said before,
$\Sigma (\omega)
\propto \omega^{2/3}$ and $m^*/m_b = Z^{-1}
\propto
 \omega^{-1/3}$ at a 2D
 nematic QCP.  The transport  scattering rate $\gamma_{\text{tr}}$ is obtained by multiplying a single-particle scattering rate ($\Sigma^{''} \
 \propto \omega^{2/3}$)  by a ``transport factor" $
 1 - \cos \theta
  \sim \theta^2
 \propto
 q^2_{\parallel}$, where $q_\parallel$ is a typical momentum
 transfer
  along the FS,
  which
   scales as $
  \omega^{1/3}$. Hence $\gamma_{\text{tr}}
  \propto \omega^{4/3}$.
   Substituting this result along with  $Z\propto \omega^{1/3}$
 into Eq. (\ref
 {ch_1_1})
at $\omega=\Omega$, we find that
$\sigma' (\Omega)\to\text{const}$
 at $\Omega \to 0$.  On the other hand,
 Refs.~\onlinecite{kim:1994} and \onlinecite{eberlein:2016}
find that
 $\sigma' (\Omega)
\propto 1/\Omega^{2/3}$, which obviously contradicts Eq. (\ref{ch_1_1}).

   We argue below that
   additional
   vertex  renormalization cancels out the $Z^2$ factor in Eq. (\ref{ch_1_1}), such that the modified
   version
    of Eq. (\ref{ch_1_1})
    becomes
    \beq
     \sigma' (\Omega)
     = \frac{\Omega^2_p}{4\pi}  \frac{\gamma_{\mathrm{tr}} (\Omega)}{\Omega^2}.
    \label{ch_1_2}
    \eeq
   The cancelation between the vertices and $Z$-factors is consistent with
   the
    argument \cite{kim:private,kim:1995} that the conductivity is a gauge-invariant object and, as such,
 cannot contain a $Z$-factor.  The frequency dependence of $\sigma' (\Omega) \propto \Omega^{4/3}/\Omega^2
 = 1/\Omega^{2/3}$,
  predicted by Eq.~(\ref{ch_1_2}),
 agrees with the results of Refs.~\onlinecite{kim:1994,eberlein:2016}.
   However, our calculation goes beyond the two-loop order considered in Ref.~\onlinecite{kim:1994} in that we compute the conductivity using fully renormalized Green's functions and summing up infinite series of vertex renormalizations.

   We note in passing that Eq.~(\ref{ch_1}) with $\gamma_\text{tr}=2\Sigma''(\Omega)$ can also be viewed as the result of the Kubo formula in the $D\to\infty$ limit,
   in which vertex corrections are absent.\cite{georges:1996} However, there is no contradiction with the results described above, because the $Z$-factor for a nematic QCP in $D$-dimensions behaves as $1/Z-1\propto \omega^{(D-3)/3}$, i.e, $Z\to 1$ as $\omega\to 0$ already for $D>3$. Therefore,
     there is no additional 
   singular $\Omega$-dependence of the conductivity coming from the $Z$-factor in the large-$D$ limit. $D=3$ is a marginal dimensionality, in which the $Z$-factor vanishes logarithmically, but this vanishing is also compensated by a logarithmically divergent current vertex.

We also consider a SDW criticality and analyze the contribution to the conductivity
from fermions both near hot spots and in ``lukewarm'' regions, \cite{hartnoll:2011,chubukov:2014} which lie in between hot and cold parts of the FS.
 For hot fermions we find that,  in contrast to the nematic case,
  there is no cancellation between the $Z$-factors and current vertices.  This implies that the correct result is reproduced by the extended Drude formula in Eq.~(\ref{ch_1}), which does takes mass renormalization into account. For lukewarm fermions, however, we find that there is again a cancellation between the $Z$-factors and  current vertices, which implies that Eq.~(\ref{ch_1}) breaks down. This cancellation leads to $1/\Omega$ scaling of $\sigma'(\Omega)$ for all frequencies of interest rather than at only higher frequencies, as it was argued in previous papers.\cite{hartnoll:2011,chubukov:2014}

The rest of the paper is organized as follows. In Sec.~\ref{sec:nem} we consider a
nematic
QCP. In Sec.~\ref{sec:nem_gen} we formulate the diagrammatic approach to the optical conductivity 
based on the idea of energy-scale separation.
In Sec.~\ref{sec:nem_FL} we calculate the optical conductivity in the FL region near
 to
but away
from a
nematic
QCP. In Sec.~\ref{sec:nem_qcp} we extend the analysis
right
 to the QCP. In Sec.~\ref{sec:SDW}  we consider a SDW QCP. Contribution to the optical conductivity from hot  and lukewarm fermions are discussed in Secs.~\ref{sec:hot} and ~\ref{sec:warm}, correspondingly.

\section{Nematic quantum critical point}
\label{sec:nem}
\subsection{General reasoning}
\label{sec:nem_gen}
 We consider a system of fermions on a 2D lattice  near a $T=0$ Pomeranchuk-type transition into a state which breaks lattice rotational symmetry.
 [Alternatively, one can consider a ferromagnetic QCP, provided that the continuous quantum phase transition is stabilized by lowering the spin symmetry
 from $O(3)$ to $Z_2$,\cite{rech:2006} or else a model of fermions coupled to $U(1)$ gauge field.\cite{kim:1994}]
 We assume, as in earlier studies,  that near
 the
  transition the effective
electron-electron interaction is mediated by the dynamical susceptibility of the order-parameter field
\beq\label{chi}
 \chi (q, \Omega_m) = \frac{\chi_0}{q^2 + M^2 + \gamma |\Omega_m|/q},
 \eeq
 where $M$ is the inverse correlation length of order-parameter fluctuations
 (bosonic mass).
 We assume that
 the
 fermion-boson coupling is
 $g f(\bk)$, where
  $g$ is a constant roughly of order Hubbard $U$, ${\bf k \pm \bf q/2}$ are momenta of fermions that couple to a boson with momentum $\bf q$, and $f(\bk)$ is the form-factor associated with the rotational symmetry of the order-parameter field.   The effective coupling, which appears in the formulas below
 for the fermionic self-energy and conductivity, is   ${\bar g} (\bk) = g^2 f^2 (\bk) \chi_0$. The factor $f(\bk)$ will not play any significant role in our analysis
 and, to simplify the presentation, we neglect the ${\bf k}$ dependence of ${\bar g}$.

 We begin by listing 
 the
 known facts about the system behavior near a nematic
 QCP.
  The notations are simplified by assuming that the Fermi system is isotropic, which is what we will do in what follows. Anisotropy can be readily restored but it will not be necessary.
  First, the Landau damping term in the bosonic propagator comes from the same fermion-boson interaction, and the prefactor $\gamma$ of this term scales as $\gamma \sim {\bar g} k_F/v^2_F$,
  where $k_F$ and $v_F$ are the Fermi momentum and velocity, correspondingly.
 Second, sufficiently close to the QCP,
  i.e.,
 for $M^2 \ll
 m\bar g
 $,\cite{chubukov:2005d} the fermionic self-energy
  depends much stronger on the frequency than on
 the momentum
 and has the form
  \beq
 \Sigma (\omega_m) = i \lambda \omega_m f_\Sigma\left(\frac{|\omega_m|}{\omega_{{\rm FL}}}\right).\label{sigma}
 \eeq
 Here,
 \bea
 \lambda = {\bar g}/4\pi v_F M\label{lambda}
 \eea
 is the dimensionless coupling constant,
 \bea
 \ofl
 = M^3
 /\gamma\label{ofl}\sim \frac{M}{k_F} \frac{(Mv_F)^2}{\bar g}
\eea
  is the energy scale separating the FL and NFL regimes ($\omega\ll \ofl$ corresponds to a FL
  and {\em vice versa}),
 and  $f_{\Sigma}(x)$ interpolates between the limits of $f_{\Sigma}(x \ll 1) = 1 + {\cal O}(x)$ and $f_{\Sigma} (x \gg  1) \propto x^{-1/3}$.
  In the FL regime,  $\Sigma (\omega_m) \approx i \lambda \omega_m+i\text{sgn}\omega_m a \omega_m^2
  $, where $a\sim \lambda/\ofl$.
 The corresponding
 real-frequency Green's function  is given by
  \beq
  G(\bk,
  \omega) = \left[\omega/Z - \ve_\bk+i\Sigma''(\omega)\right]^{-1},\label{gf}
  \eeq
 where $Z = 1/(1+ \lambda)$ and $\Sigma''(\Omega)=a\omega^2$.

We next turn to the conductivity.  Because the interaction is peaked at $q=0$ (i.e., it is long-ranged in the coordinate space), umklapp scattering
is suppressed.~\cite{maslov:2011, pal:2012b}
 Therefore, the {\em dc} conductivity can be rendered finite only by impurities or non-critical channels of  the interaction.
However, for fermions on a lattice  $\sigma' (\Omega)$
is  finite even
if only normal, i.e.,
momentum-conserving,
electron-electron scattering is present.\cite{gurzhi:1959,maslov:2017} Furthermore, if the
 FS contains inflection points, as we assume to hold in our case,
    the conductivity due to normal scattering is not reduced compared to what one would get if
    normal and umklapp scatterings were
    comparable.
    \cite{gurzhi:1982, rosch:2005, rosch:2006, maslov:2011, pal:2012b, briskot:2015}

\begin{figure}[h]
 \includegraphics[width=1.0 \linewidth]{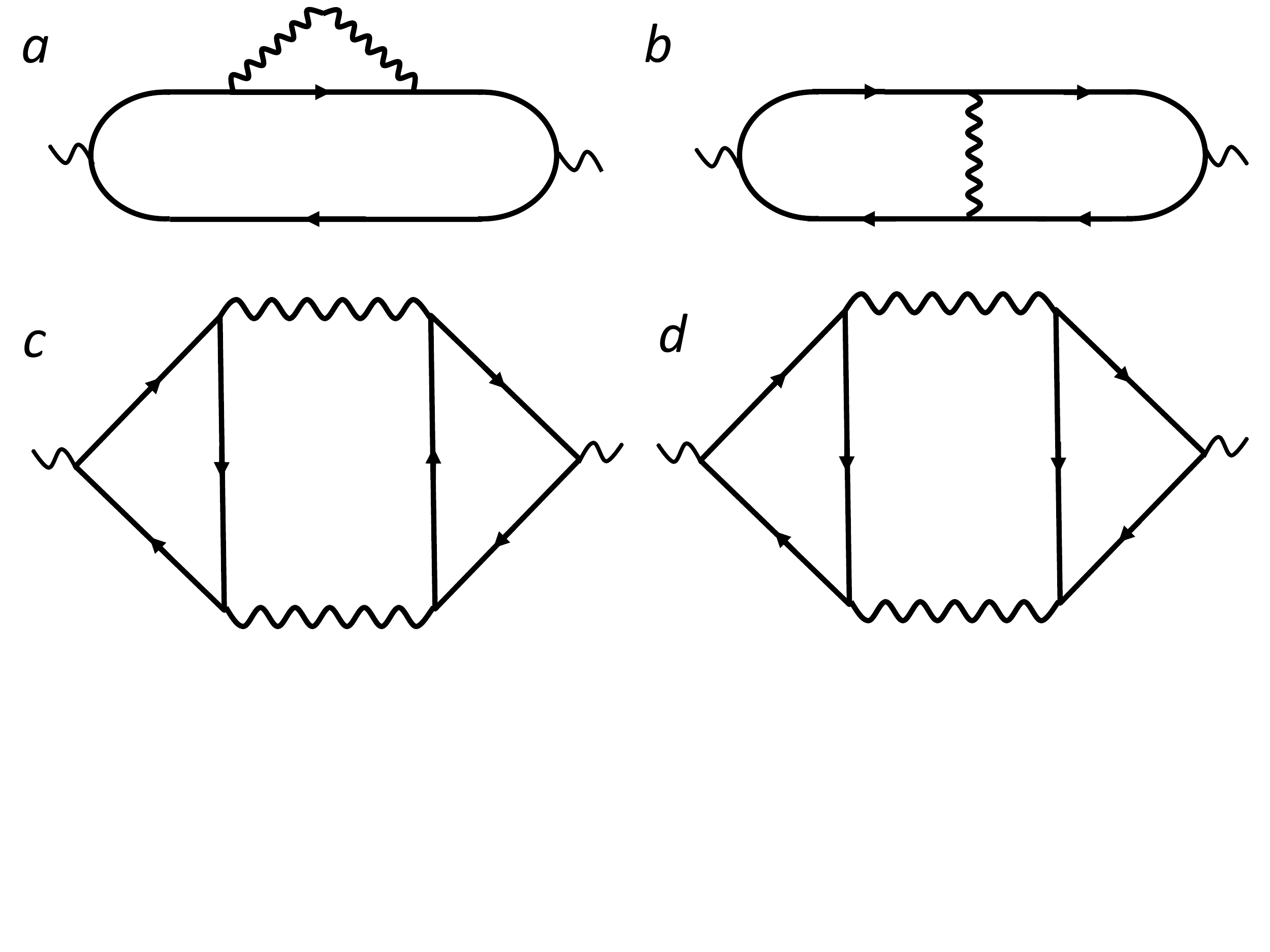}
\vspace{-0.9in}
\caption{Maki-Thompson ({\em a-b}) and Aslamazov-Larkin ({\em c-d}) diagrams for the conductivity. The mirror image of diagram {\em a} is not shown.
This solid lines denote bare Green's functions.
 \label{fig:MTAL}}
\end{figure}

 The most straightforward way to obtain  $\sigma' (\Omega)$ is to use
 the Kubo formula, which relates  $\sigma' (\Omega)$  to
 the imaginary part of the current-current correlation function at $q=0$, $K'' (\Omega)$:
 \beq
  \sigma' (\Omega) =
\frac{K'' (\Omega)}{\Omega},
\label{1}
\eeq
In the diagrammatic representation, the current-current correlator $K(\Omega)$ is
a
 fully dressed particle-hole bubble with  current vertices on both sides.  To the lowest order in ${\bar g}$, the imaginary part of $K$ comes from four diagrams shown in
Fig.~\ref{fig:MTAL}.
We will be referring to
diagrams {\em a-b}
as to Maki-Thompson diagrams,  and to diagrams {\em c-d}
as to
Aslamazov-Larkin ones. The latter are actually of the same order as the Maki-Thompson diagrams, despite that
they formally contain
 an
extra power of ${\bar g}$.
\cite{holstein:1964,riseborough:1983,yamada:1986,gornyi:2004,maslov:2017}.  The reason, in our case,  is that the contribution to  $K'' (\Omega)$
  from
  the
  Maki-Thompson diagrams  comes from the dynamical part of the bosonic propagator -- the Landau damping
  term.
 The latter appears in the bosonic propagator due to coupling to fermions and  contains ${\bar g}$ in the prefactor.   This makes
 the
 Maki-Thompson contribution to $K'' (\Omega)$ of the same order as
 the
 Aslamazov-Larkin one.

For a Galilean-invariant system,
momentum conservation implies current conservation and thus $\sigma'(\Omega)$ must vanish.
Consequently,
the Maki-Thompson and Aslamazov-Larkin contributions to $K'' (\Omega)$ cancel each other.\cite{riseborough:1983,yamada:1986,gornyi:2004,maslov:2017} For fermions on a lattice (our case)
 momentum conservation does not imply current conservation and $\sigma'(\Omega)$ does not have to vanish.
In this case,  the Maki-Thompson and Aslamazov-Larkin contributions
are
 generically
of the same order,
 but do not cancel each other.  To obtain the frequency dependence of $K''$  it is then sufficient to consider only one of these contributions
 and return to the isotropic case. The actual result will differ from the one obtained under these approximations
 only
  by a factor of order one,
 which reflects anisotropy of the Fermi surface.
In what follows,
 we will focus on
the
 Maki-Thompson diagrams.

\begin{figure}[h]
\includegraphics[width=1.0 \linewidth]{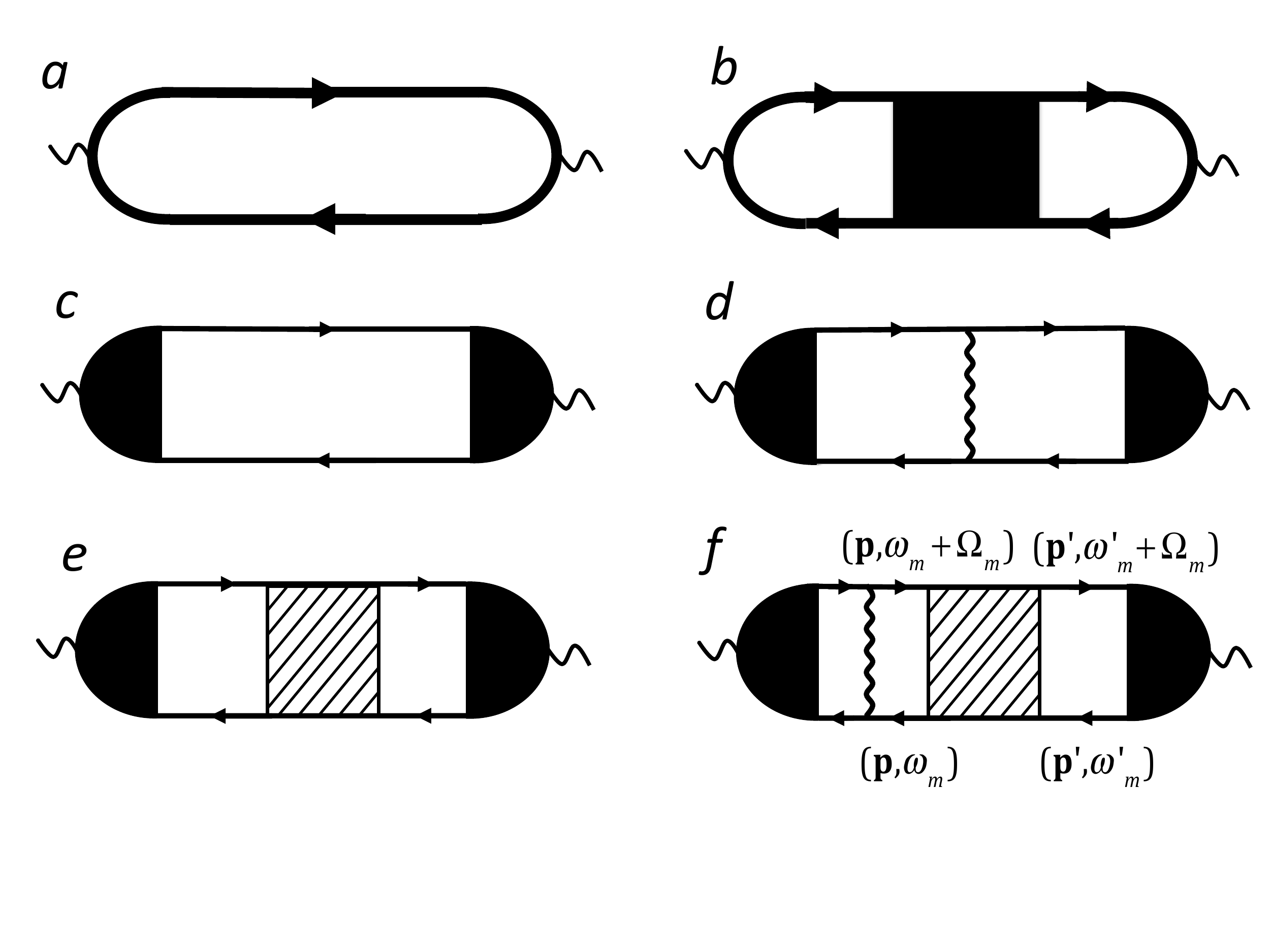}
\vspace{-0.5in}
\caption{Separation of energy scales for the conductivity. The sum of diagrams {\em a} and {\em b} represent an exact current-current correlation function. Exact Green's functions are denoted by thick lines. Diagram {\em c} consists of two low-energy Green's functions given by Eq.~(\ref{gf}) (thin lines) and current vertices, which include all high-energy renormalizations. Diagram {\em d} provides a vertex (transport) correction to diagram {\em c}.
Diagrams {\em e} and {\em f} vanish for zero external momentum and finite frequency, which is the case for the conductivity.
 \label{fig:side}}
\end{figure}

The fully renormalized
 current-current correlator is shown in Fig.~\ref{fig:side}.  It is expressed in terms of fully dressed fermionic Green's functions and
a fully dressed
 four-leg
 vertex. At
 small ${\bar g}$ and away from the critical point
 (but still such that $M\ll k_F$),
 the dimensionless coupling $\lambda$ is small. Then the lowest-order approximation in
the fermion-boson coupling is sufficient and
 we go back to diagrams  {\em a-b} in Fig. \ref{fig:MTAL}, where the solid lines are the propagators of free fermions. Evaluating these diagrams  we obtain the known FL result\cite{gurzhi:1959}
\bea
\sigma'(\Omega)=\frac{\Omega_p^2}{4\pi} \frac{
\gamma_{\text{tr}} (\Omega)}{
\Omega^2},
\label{sp_1_1}
\eea
where
the
 transport
 scattering rate
  $\gamma_{\text{tr}} (\Omega)
  \sim
  (M/k_F)
  ^2 \Sigma''(\Omega)$ and $
 \Sigma''(\Omega) \sim
 \lambda\Omega^2/\ofl\propto \Omega^2/M^4$ is the single-particle
 scattering rate.
  The smallness of $\gamma_{\text{tr}}$ compared to
  $
  \Sigma''
  $  reflects the fact that small-angle
  scattering is inefficient for momentum relaxation.  Mathematically, the factor of $(M/k_F)^2$ in $\gamma_{\text{tr}}$ appears because the two Maki-Thompson diagrams partially
 compensate  each other.
 Substituting $\gamma_{\text{tr}}$ into Eq.~(\ref{sp_1_1}), we find that the conductivity does not depend on 
 frequency
and scales with $M$ as
 \bea
\sigma'(\Omega)
\propto  M^{-2}.\label{sp_1_2}
\eea
This is a familiar  ``FL foot'': a plateau in the frequency dependence
of the optical conductivity of  a FL.\cite{gurzhi:1959,berthod:2013}

A naive way to go beyond the lowest order would be to replace the bare Green's functions in diagrams {\em a} and {\em b} in Fig.~\ref{fig:MTAL} by the renormalized ones, which contain the self-energies in the denominators. The real part of the self-energy would then renormalize the external frequency by a factor of $1+\lambda=1/Z$. Expanding diagram {\em a}  in the imaginary part of the self-energy, we would then get
 instead of  Eq.~(\ref{sp_1_2})
 \bea
\sigma'(\Omega)=\frac{\Omega_p^2}{4\pi} \frac{Z^2 \gamma_{\text{tr}} (\Omega)}{\Omega^2} \propto \frac{Z^2
}{M^2}. \,\label{sp_4}
\eea

If this result could be extended to a strong-coupling limit, where  $Z\propto M^2$,
we would arrive at the conductivity that is independent of $M$ in the limit of $M\to 0$. Since $M$ drops out,
we would then conclude that $\sigma' (\Omega)$ remains to be constant  even right at the QCP, where $M=0$.
 However, it is obvious that the method described in the preceding paragraph is not consistent even in the weak-coupling limit, where $\lambda\ll 1$ and $Z\approx 1$. Indeed, recalling that $\Sigma''\propto \bar g$ and $Z=1+{\cal O}(\bar g)$, we see that dependence of $\sigma'(\Omega)$ in Eq.~(\ref{sp_4}) on the coupling constant is
\bea
\sigma'(\Omega)\propto \frac{\bar g}{[1+{\cal O}(\bar g)]^2}\sim \bar g +{\cal O}(\bar g^2)+\dots
\eea
Taking into account the effect of mass renormalization (the denominator in the equation above) amounts to finding a second-order correction to the conductivity
in the coupling constant. This means that all second-order vertex corrections also need to be collected, but we accounted only for those which are obtained by inserting self-energy corrections into diagram {\em b} in Fig.~\ref{fig:MTAL}.

Collecting corrections to the current vertex is simplified in our case of a long-range interaction,
 because the current vertex $\bar{\bf \Gamma}$
for an incoming fermion with momentum $\bk$
 is related to the density vertex, $\Gamma$, simply by $\boldsymbol{\bar\Gamma}={\bf v}_
{\bk}\Gamma$, up to
small
corrections (here, $\bv_\bk=\partial_\bk\ve_\bk$).
If the momentum carried by the wavy line in diagram {\em d} in Fig.~\ref{fig:side} is $\bq$, then the left current vertex in this diagram is replaced
by $\bv_\bk\Gamma$ and the right one by $\bv_{\bk+\bq}\Gamma$ with $k=k_F$.\footnote{It can be shown that keeping terms of order $q$ in the integral equation for the current vertex gives corrections of order $\Omega/E_F$, which can be safely discarded.} On other hand, the current vertices in diagram {\em b} give $\bv_\bk^2\Gamma^2$. The combination $\bv_\bk^2-\bv_{\bk}\cdot\bv_{\bk+\bq}$ gives the transport factor, which we discussed above, and now the problem
reduces to finding the renormalized charge vertex $\Gamma$.

The strength of the renormalization of  $\Gamma$ depends on the
 ratio of the
 external
  momentum and frequency. We are interested in the regime where the
  external
   momentum is zero, while
  the external frequency ($\Omega$) is finite.
  [The opposite limit is discussed in Appendix \ref{sec:appA}.]
  In this regime, the density vertex
  satisfies
  the
   Ward identity
   following from the particle number conservation:
$\Gamma (\Omega) = 1 +  \left[\Sigma (\Omega+ \omega) - \Sigma (\omega)\right]/\omega$.  Using $\Sigma (\omega) = \lambda \omega$, we immediately obtain $\Gamma = 1 + \lambda = 1/Z$.

Nevertheless,
 inserting vertex corrections into the formula for conductivity is
 still
a tricky issue because
 diagram {\em b} in Fig.~\ref{fig:MTAL}, which we already included into
Eq.~(\ref{sp_1_1}), is
 also
 a vertex correction.
   This contribution and the ones that
  renormalize the vertex in accord with the Ward identity can be separated if
 one assumes that they come from different
  energy scales.
  The idea of energy scale separation in a FL (which is similar to the underlying idea of renormalization group) was put forward by Eliashberg in the context of {\em dc} conductivity  \cite{eliashberg:1962}
 and has been used to calculate various correlation functions of both clean \cite{ipatova:1962,larkin:1972,shekhter:2005,chubukov:2014b} and dirty \cite{finkelshtein:1983, *finkelshtein:2010} FLs. In real-time formulation, this method  amounts to representing a diagram  for any given  correlation function by a sequence of irreducible vertices separated by pairs of low-energy retarded (R) and advanced (A) Green's functions given by Eq.~(\ref{gf})  (``RA sections'').\cite{finkelshtein:1983, *finkelshtein:2010} Because the method neglects diagrams with
 crossed irreducible vertices, it is equivalent to a kinetic equation for a FL, which
 takes into account the residual interaction between quasiparticles via an appropriate collision integral.\cite{eliashberg:1962}

 Any system that exhibits a FL behavior does so only at energies below certain scale which, in general, is smaller than the Fermi energy and is determined
 by the dynamics of the effective interaction.
 In our case, such ``high-energy'' scale is $\ofl$ given by Eq.~(\ref{ofl}). The second, ``low-energy'' scale is determined by energies  which the system
 is probed at. In our case, the external frequency ($\Omega$) plays the role of such a scale.

 As long as $M\neq 0$, $\ofl$ is finite, and we can
 choose $\Omega$ to be smaller than $\ofl$.
 In this situation,
   one can
   evaluate
   the
    diagrams for the conductivity
by using the separation-of-scales method.
 Namely,
  one
  selects
 a  cross-section containing a pair of low-energy Green's functions
  [Eq.~(\ref{gf})] in 
  a
  diagram of 
   arbitrary order, as shown in Fig.~\ref{fig:side}{\em c}.
All other elements  of the diagram to the left and right of this cross-section are combined  into
two
renormalized
 side vertices.
Next, one
selects a cross-section composed of four low-energy Green's function intersected by a wavy line and again lumps the rest of the diagram
into the left and right vertices, as shown in  Fig.~\ref{fig:side}{\em d}. Then
 one
 expands the low-energy Green's functions in diagram {\em c}
 to first order in $\Sigma''$ and
   neglects
   $\Sigma''$ in the Green's functions forming the central part of diagram {\em b}.
 As a result,
 one gets diagrams
 of the same structure of
 as in
 Fig. \ref{fig:MTAL} {\em a} and {\em b}, but with renormalized side vertices.~\cite{eliashberg:1962}
 The sum
 of the central parts
   of
  diagrams {\em c} and {\em d} in Fig.~\ref{fig:side} yields Eq.~(\ref{sp_1_1}),
 while
  the renormalized side vertices give two
  factors of $\Gamma$.  The full answer then becomes
  \bea
\sigma'(\Omega)=\frac{\Omega_p^2}{4\pi} \frac{Z^2 \Gamma^2 \gamma_{\text{tr}} (\Omega)}{\Omega^2} \propto \frac{Z^2 \Gamma^2}
{M^2}  =  \frac{1
}{M^2}. \label{sp_1_3}
\eea
In the last equation we used $\Gamma = 1/Z$.
This is the same result as obtained in the weak-coupling limit [Eq.~(\ref{sp_1_2})]:
 the conductivity
tends to  a finite value proportional to $1/M^2$ at
 low frequencies.  At $M \to 0$,
 $\sigma' (\Omega)$
formally diverges, but near a critical point
 FL
regime
extends
only
up to $\Omega\sim \ofl\propto M^3$.  At
higher
 frequencies, one can use standard scaling arguments and replace $M$ by $\Omega^{1/3}$.
This yields $\sigma' (\Omega) \propto 1/\Omega^{2/3}$, in agreement with the results obtained perturbatively \cite{kim:1994}  and
via dimensional regularization.\cite{eberlein:2016}

Note that if we were interested in a correlation function taken in the opposite limit, when $\Omega/v_F$ is less than the external momentum $Q$,
the ladder series must have been continued by selecting
more low-energy cross-sections, separated by irreducible vertices, as shown in diagrams {\em e} and {\em f} in Fig.~\ref{fig:side}.
An appropriate irreducible vertex for this case would be the FL vertex $\Gamma^\omega$, which is related to the Landau interaction function.\cite{agd:1963}
The resummation of  the geometric series in $\Gamma^\omega$ for, e.g., the spin susceptibility $\chi_s$, is necessary to reproduce
the denominator in the FL result for $\chi_s$ in the limit of $\Omega=0$ and $Q\to 0$: $\chi_s=N_F^*/(1+F^a_0)$, where $N_F^*$ is the renormalized density of states.\cite{chubukov:2014b} However, the conductivity is
 obtained
 at finite $\Omega$ and  $Q=0$. In this case diagrams {\em e}, {\em f}, and similar diagrams of higher orders vanish. To see this,
we label the incoming and outgoing states of the irreducible vertex (hatched box) as shown in diagram {\em f}. Since the irreducible vertex is
as a high-energy object of the theory, one can safely neglect its dependence on low-energy  variables, i.e., $\ve_\bp$, $\omega_m$, etc.,
and 
consider it to be a function  only of the angle between $\bp$ and $\bp'$. Then the integral over $\omega_m'$ of the two Green's functions to the right of the hatched box
 vanishes because the poles of the integrand are located in the same half-plane.
 We thus
conclude that diagrams {\em c} and {\em d} indeed give  the full result for the conductivity, provided that $\Omega/Z\gg \gamma_{\text{tr}}$.

 \begin{figure}[h]
\vspace{-0.1in}
\includegraphics[width=1.0 \linewidth]{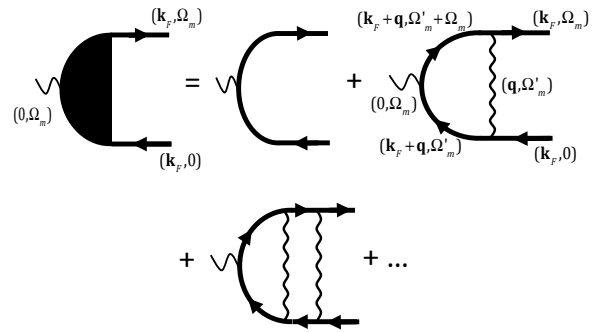}
\vspace{-0.5in}
\caption{Vertex renormalization  for a nematic QCP.  Thick solid lines denote exact Green's functions.
The wavy line is the susceptibility from Eq.~(\ref{chi}). \label{fig:vertex}}
\end{figure}

The  issue that we address in this paper is whether it is indeed possible to separate
two types of
contributions to the
optical conductivity:
the one which determines the transport scattering rate $\gamma_{\text{tr}}$
and the one
which
accounts for
renormalization
 of the current vertices.   We
 argue
 that this is a non-trivial issue even in the
 FL  regime, where we do have two different scales: $\Omega$ and  $\ofl$. We show that the contributions from
 diagrams {\em a} and {\em b} in Fig.~\ref{fig:MTAL},
 which add up to $\gamma_{\text{tr}}$,
  come from internal
 frequencies of order $\Omega$, and this holds
 regardless of whether
 one integrates first over
 the
  internal frequency or over
  the
  fermionic dispersion $\ve_\bk$.  The issue
  of vertex renormalization
  is more subtle.
     The
     renormalized
  current vertex (which, we remind, in our case is the same as the density vertex multiplied by the Fermi velocity) is obtained by summing up
  the ladder series
  shown in Fig.~\ref{fig:vertex}.
Non-ladder diagrams are smaller at each given other.
The building block
($B$)
of the
 ladder series is the convolution of two fermionic propagators and one bosonic propagator;
 symbolically,
 $B=\int G G \chi$.
  The
  double
  integral over
 the
 internal frequency and fermionic dispersion is convergent, hence the result does not depend on the order of integration.  Yet,
 characteristic internal energies, encountered when integrating in different order, differ.
 The fastest
 way to evaluate the
 $B$
 is to
  integrate
  over $\ve_\bk$ first. Then the
  integral
  is determined by
  the poles
  of
  the fermionic propagators (see Sec.~\ref{sec:vertex} below). This calculation gives $B=\lambda/(1+ \lambda)$.
  The ladder series
  of $B$ blocks
  is
    geometric,
  so the full result for the vertex is  $\Gamma = 1/(1-
  B) = 1+\lambda$,
  in agreement
  with the Ward identity.

The problem with
applying
this approach
to
 the conductivity
 is that
 typical internal
 frequencies
 and 
 fermionic dispersions
 in
 $B$ are of order $\Omega$, i.e.,
 comparable to
 the
 characteristic frequency
 that
  determine $\gamma_{\text{tr}}$.  In this situation, one cannot separate
  a
  computation of vertex corrections from
  that
   of $\gamma_{\text{tr}}$.  This poses a real problem because at  large $\lambda$ the building block
   of
   the ladder series $B=\lambda/(1+\lambda)$
    is approximately
   equal to
   unity,
  and the sum of such terms converges for any finite $\lambda$ only because the numerical prefactors of all terms are equal to unity as well,
  i.e., the series is geometric.
      If one cannot separate
      the energy
      scales, then each term in the ladder series gets multiplied by a factor of order $Z^2\gamma_{\text{tr}}/\Omega$ from the internal part of the diagram, but the
      numerical
      coefficients
      now do not necessary correspond to a geometric series, and  the new series is not guaranteed to converge for any $\lambda<\infty$.
     It is also not guaranteed that the answer will contain $\Gamma^2$, i.e., that
     each
     side vertex
      in the current-current correlator gets renormalized by $1+\lambda$.   The situation is even worse in the
      NFL regime, i.e.,
      for $\Omega\gg\ofl$,
     where
       $\Sigma' \sim \Sigma'' \sim  \omega^{1/3}_0\omega^{2/3}$.  The energy
       \bea
       \omega_0=\bar g^2/v_Fk_F\label{om0}
       \eea separates the perturbative regime, where $\omega\gg\omega_0$
       and hence $|\Sigma|\ll \omega$, from the non-perturbative one, where $\omega\ll\omega_0$ and hence $|\Sigma|\gg \omega$.
       In the non-perturbative region, i.e., for external $\Omega\ll\omega_0$,
             each
      term in the ladder series for the vertex is
          a number of order one, and the series
      is not geometric.
           The solution of the integral equation for the vertex shows
     that  ladder series is summed into \cite{chubukov:2005c}
       $\Gamma (\Omega)
       \approx
       (\omega_0/\Omega)^{1/3}
       \approx 1/Z(\Omega)$, in agreement with the  Ward identity.
       However, this
       relation
        holds
       due to
       specific ratios
       of ${\cal O}(1)$ terms
       at
       consecutive
        orders.
        Once one combines vertex renormalization at a given order with the part of the diagram that gives
        $Z^2 \gamma_{\text{tr}}/\Omega$,
        the ratios
        of
        terms at consecutive orders
         change, and there is no guarantee that the sum of ladder series will be of order $1/\Omega^{1/3}$.
         In addition,
         without a separation of scales  there is
         no argument for why the diagrams for the current-current correlator
         should contain two renormalized current vertices rather than one.

We show in Sec.~\ref{sec:nem_FL}  that
the separation of
 scales is actually possible in the
 FL regime but,
to apply this method in a consistent manner,
one should evaluate the building block for the vertex correction
in
 a different order:
by integrating first over
the
fermionic frequency and then over $\ve_\bk$. This way, the integral over frequency comes from the branch cut in the bosonic propagator.
In the Matsubara representation,
 this branch cut is associated with
 a non-analytic, $|\Omega_m|$ frequency dependence of the Landau damping
  term.
  In this computational scheme, typical internal $\omega$ and $\ve_\bk$ in $B=\int GG \chi$ are of order $\ofl$ rather than
  $\Omega$.  Then the separation of
  scales is possible
  as long as $\Omega
  \ll  \ofl$. As
  a consequence, the conductivity $\sigma' (\Omega)$ has the form of Eq.~(\ref{sp_1_3}).
   In the NFL regime,
  the separation of scales is not, strictly speaking,
  possible.
  Still, in  Sec.~\ref{sec:nem_qcp} we present a  renormalization group-type argument
  which shows
  that
  the $1/M^2$ dependence of $\sigma'(\Omega)$ at $M\neq 0$ translates into $\sigma'(\Omega)\propto \Omega^{-2/3}$ in the NFL regime.

We note in passing that
   there exists
 another energy scale in the NFL regime:
 $ \omega_0$   defined
  in Eq.~(\ref{om0}).
 However, we will show below that there is no contribution to the vertex correction from this scale, no matter in what order
 the integrals in $B$
 are
 evaluated.

\subsection{Fermi-liquid regime\\ near a nematic quantum critical point}
\label{sec:nem_FL}

 \subsubsection{Central parts of diagrams for the current-current correlation function}
 We first analyze
 diagrams
  {\em a}  and {\em b} in Fig.~\ref{fig:MTAL} and show that they are determined by
   low-energy fermions, with frequencies
of
 order of $\Omega$,
 regardless of the
 order
 in which the integrals over internal
 frequencies and
 fermionic dispersions
  are evaluated. For definiteness and for future comparison with the calculation of the
renormalized vertex,
   we integrate over frequency first and then over $\ve_\bk$.

We expect
 diagrams  {\em a}  and {\em b}  to produce the FL result $\sigma'(\Omega) \propto Z^2\gamma_{\text{tr}}(\Omega)/\Omega
^2
=\text{const}$.
 According to Eq.~(\ref{1}), the $\Omega$-independent $\sigma'(\Omega)$
 implies that $K''(\Omega)\propto \Omega$. The linear-in-$\Omega$ part of $K(\Omega)$ can be calculated directly
on the Matsubara axis; we only have to subtract  the static part of the effective interaction [Eq.~(\ref{chi})], because static
interaction
  does not give rise to
  damping of quasiparticles
  and thus does not yield
  $K''$.  The combined
  contribution to
  $K(\Omega_m)$ from  diagrams {\em a} and {\em b} reads
\bwt
\bea
K(\Omega_m)=
e^2
\int\frac{d^2q}{(2\pi)^2} \int \frac{d\Omega'_m}{2\pi }\int \frac{d^2k}{(2\pi)^2}\int \frac{d\omega_m}{2\pi}  && \frac{ \left({\bf v}_{\bk} - {\bf v}_{{\bf k} + {\bf q}}\right)^2\chi_{\text{dyn}}(q,\Omega_m' )}{\left[\frac{i\omega_m}{Z}-\ve_\bk\right]\left[\frac{i\left(\omega_m+\Omega'_m\right)}{Z}-\ve_{\bk+\bq}\right]\left[\frac{i\Omega'_m}{Z}-\ve_{\bk+\bq}+\ve_{\bk}\right]\left[\frac{i(\Omega'_m+\Omega_m)}{Z}-\ve_{\bk+\bq}+\ve_{\bk}\right]},\nn\\
,\label{K}
\eea
\ewt
where $\chi_{\text{dyn}}(q,\Omega_m')=\chi(q,\Omega_m')-\chi(q,0)$.
The factor of $({\bf v}_{\bk} - {\bf v}_{{\bf k} + {\bf q}})^2 \propto q^2$ appears when we sum up diagrams {\em a} and {\em b}. Physically, it
  accounts for the difference between the transport and single-particle scattering rates. Since typical $q
  \ll k_F$, we approximate $\ve_{\bk+\bq}-\ve_\bk$ by $v_Fq\cos\theta$, and integrate the product of the first two factors
 in
 Eq.~(\ref{K}) first over $\omega_m$ and then over $\ve_\bk$. This gives
 \bwt
\bea
K(\Omega_m)\propto Z \int dq q^4\int d\Omega'_m \int \frac{d\theta}{2\pi}
\frac{ \cos\theta}{\left[\frac{i\Omega'_m}{Z}-v_Fq\cos\theta\right]^2\left[\frac{i(\Omega'_m+\Omega_m)}{Z}-v_Fq\cos\theta\right]\phantom{^2}}\chi_{\text{dyn}}(q,\Omega_m')
\label{K2}
\eea
\ewt
We assume and then verify that typical internal frequencies $\Omega_m'$ are of order  $\Omega_m$ and typical $q$ are
of order $M$. For $\Omega'_m\sim
\Omega_m\ll v_F q\sim v_F M$, the angular integral in Eq.~(\ref{K2}) is reduced to
\bea
\int \frac{d\theta}{2\pi}\dots=\frac{Z}{\left(\Omega_mv_Fq\right)^2}\left(|\Omega'_m|-|\Omega_m'+\Omega_m|+\text{sgn}\Omega'_m \Omega_m\right).\nn\\
\label{K3}
\eea
In the same limit,  $\chi_{\text{dyn}} (q,\Omega_m')\approx-\chi_0\gamma|\Omega_m'|/q(q^2+M^2)^2$.
Substituting this into Eq.~(\ref{K2}), we obtain
 \bea
K(\Omega_m)&\propto& Z^2 \int \frac{dq q}{(q^2 + M^2)^2}\label{K2_1}\\ &\times& \int \frac{d\Omega'_m}{\Omega^2_m}
|\Omega_m'| \left(|\Omega'_m|-|\Omega_m'+\Omega_m|+\text{sgn}\Omega'_m \Omega_m\right).\nn
\eea
 As expected, the integral over $q$ is determined by $q\sim M$ and gives
 a factor of $
1/M^2
$.  The frequency integral, on the other hand, is confined to
the region 
$0\leq|\Omega'_m|
\leq |\Omega_m|$ and gives
a factor of
$
\Omega_m$.  Continuing analytically from $\Omega_m$ to real $\Omega$, we obtain
\beq
K''(\Omega)\propto \frac{Z^2\Omega}{M^2}.\label{7}
\eeq

Alternatively, one can compute the integrals in Eq.~(\ref{K}) in a different way, but splitting $\bq$ into the
components
 tangential ($q_{||}$)  and normal ($q_\perp$) to the FS.
 Integrating over $\omega_m$, $\bk$, and $q_{||}$, we obtain
  \bea
  K (\Omega_m) &\propto& \frac{
  Z
  }
  {M^2} \int d q_\perp  q_\perp   \label{6}\\&\times& \int d \Omega^{'}_m \frac{|\Omega^{'}_m|}{\left[\frac{i\Omega^{'}_m}{Z} - v_Fq_\perp\right]^2 \left[
  \frac{i(\Omega^{'}_m + \Omega_m)}{Z} -v_Fq_\perp\right]}.\nn
  \eea
For a given sign of $q_\perp$, we now integrate over that
half-plane of complex $\Omega^{'}_m$ which does not contain poles, and choose the contour to avoid the branch cut along the imaginary axis, where $|\Omega^{'}_m| = \pm iz$.  Combining then the contributions from positive and negative $q_\perp$ and  rescaling $q_\perp = x \Omega_m$, $z = y \Omega_m$,
we reduce the double integral in Eq.~(\ref{6}) to
 \bea
 K (\Omega_m) &\propto& \frac{
  Z^2
  }
  {M^2} \Omega_m \int_0^\infty d x \int_0^\infty dy \frac{xy}{(x+y)^3 \left[(x+y)^2 + 1\right]} \nonumber \\
  & =&\frac{\pi Z^2 \Omega_m}{12 M^2}.
 \eea
Continuing analytically to real frequencies,
 we reproduce  Eq.~(\ref{7}).

 Equation (\ref{7}) is the expected FL result:  $K'' (\Omega) \propto Z^2 \gamma_{\text{tr}}/\Omega$, where $\gamma_{\text{tr}} \sim \Omega^2/M^2$, i.e.,  $
 \gamma_{\text{tr}}(\Omega)
 \propto
  M^2
  \Sigma'' (\Omega)$.  For our purpose, the key element
 of this result is that  the integrals over the internal $\ve_\bk$ and $\Omega_m'$
  come from the regions confined by the external frequency $\Omega$.

We now check what are typical internal $\ve_\bk$ and $\Omega_m'$ in the vertex correction diagrams.

\subsubsection{Vertex renormalizaton}
\label{sec:vertex}

 The diagrammatic series for the
 charge vertex $\Gamma$ at zero external
 momentum and finite external
 frequency $\Omega_m$ is shown in Fig.~\ref{fig:vertex}.
   The fermionic Green's functions in
   this
    series are the full ones: $G(\bk, \omega_m) = \left[i \omega_m + \Sigma (\omega_m) - \ve_{\bk}\right]^{-1}$.  We remind that
    this form reduces to
    \bea
    G(\bk, \omega_m) = \left[i \omega_m (1 + \lambda) - \ve_{\bk}\right]^{-1}\label{gfm}
    \eea
    for $\omega_m  \ll\ofl$.
    As in the previous section, we
    assume that $T=0$, in which case bosonic and fermionic Matsubara frequencies are continuous variables.  For simplicity, we  set the frequency of an incoming fermion to be zero
    (then the frequency of an
  outgoing fermion
  is $\Omega_m$), and set
  the
  fermionic momentum
 (which is the same for incoming and outgoing fermions) to be $\bk_F\equiv k_F\bk/k$. The series reads
  \beq
  \Gamma=
  1 +
  \Gamma_1+ ...,
  \eeq
  where
  \bea
\Gamma_1 &=&
 g^2
\int  \frac{d\Omega'_m}{2\pi} \int  \frac{d^2q}{(2\pi)^2}
G(\bk', \Omega'_m) G(\bk', \Omega'_m + \Omega_m)\nn\\
&&\times \chi (q,\Omega'_m)\label{gamma1}
  \eea
  and $\bk'=\bk_F+\bq$.
  We remind that $g^2$ is related to ${\bar g}$, which we used earlier, by ${\bar g} = g^2 \chi_0$.
As before,
   the bosonic momentum $\bq$ can be decomposed into the components perpendicular and tangential to the FS, $q_\perp$ and $q_{||}$, correspondingly. With this decomposition,  the fermionic dispersion in the Green's functions entering
   Eq.~(\ref{gamma1}) can be approximated as $\ve_{\bk'}=v_Fq_\perp$.

   We assume and then verify that  typical $q_\perp$ in the integral in Eq.~(\ref{gamma1}) are much smaller than typical $q_{||}$. Using this assumption, we neglect $q_{\perp}$ in $\chi(q,\Omega_m')$. Integration over $q_{||}$
  is then elementary and gives an effective local susceptibility
  \beq\chi_L (\Omega'_m) = \int \frac{dq_{||}}{2\pi} \chi (q_{||}, \Omega'_m)=\frac{\chi_0}{
2M} f_L\left(\frac{|\Omega'_m|}{\omega_{\text{FL}}}\right),
 \eeq
 where  $f_{L}(x \ll 1) = 1 + {\cal O}(x)$ and $f_{L} (x \gg  1) \propto x^{-1/3}$.

  The double integral over $\Omega'_m$ and $\ve_{\bk'}=v_Fq_\perp$ is convergent in the ultraviolet,  and thus the order of integration should not matter.
At the same time, the structure of the integrand is not symmetric with respect to
 $\Omega'_m$ and
 $\ve_{\bk'}$, and characteristic values of
 $\Omega'_m$ and $\ve_{\bk'}$,
 which
 contribute mostly to
 the integral, are not  necessary the same.

 Because the integral over $\Omega_m'$ formally extends into the regions where the low-energy form of the Green's function, Eq.~(\ref{gfm}), is not valid,
 it is tempting to integrate in Eq.~(\ref{gamma1}) over $\ve_{\bk'}$ first.
The integral over $\ve_{\bk'}$  is non-zero
 only if
 the poles of Green's functions are located in the opposite half-planes of $\ve_{\bk'}$, which implies that the internal frequencies are confined to the interval $-\Omega_m\leq \Omega'_m<0$  for $\Omega_m>0$ and to a similar interval for $\Omega_m <0$.  Using this form, we obtain in the FL regime ($\Omega_m \ll \ofl$):
  \beq
  \Gamma_1 = \frac{\lambda}{1+ \lambda} .
  \label{4}
   \eeq
  Evaluating higher-order diagrams in the same way, we find that they form a geometric series $1 + \Gamma_1 + \Gamma_1^2 + .. = 1/(1-\Gamma_1)$. Then
    the  full  vertex is
    \beq
    \Gamma = \frac{1}{1-\Gamma_1} = \frac{1}{1-\frac{\lambda}{1+\lambda}} = 1+\lambda.
    \eeq
    This in agreement with the Ward identity $\Gamma (\Omega) = 1 + \partial \Sigma (\omega_m)/\partial(i\omega_m)$.

  We see, however, that in this computational procedure the internal frequencies $\Omega_m'$
    are of the same order as the external
    one ($\Omega_m$), and typical $\ve_{\bk'}$ are of order of $(1+ \lambda) \Omega_m$.
   As we said in the previous section, this
   creates an ambiguity
   when
   the  diagrammatic series for vertex renormalization are combined with the
   central
    parts of
   diagrams
  {\em c} and {\em d} in Fig.~\ref{fig:side}, as
  these parts
  and
  vertex
  corrections
  come from the same energy interval.

   We now show that if the order of integrations over $\Omega'_m$ and $\ve_{\bk'}$ in Eq.~(\ref{gamma1}) is interchanged, the result
   remains the same, but typical
   $\Omega'_m$
   are now of
   order of $\omega_{{\rm FL}}$ rather than
   of
   $\Omega_m$. For $\Omega_m \ll \omega_{{\rm FL}}$, this
   provides
   a justification
    for
    the
   separation of scales,
   which is
   required for the validity of  Eq.~(\ref{sp_1_3}).

 Integration in Eq. (\ref{gamma1}) over frequency
  is rather
   complicated due to the presence of self-energies in the Green's functions.
   These self-energies cannot be replaced by either FL or NFL forms because, as will see, at least part of the result comes
   from the crossover region between the two forms.
    The integrand in Eq.~(\ref{gamma1}),
    viewed as a function of $\Omega'_m$, has poles
    from the Green's functions
    and
    branch cuts
    from
    both
    the bosonic propagator and
    self-energy.  For a given sign of $\ve_{\bk'}$,
    the poles
    of
    the two
    Green's functions
    are in the same half-plane of $\Omega'_m$, even if external $\Omega_m$ is non-zero.
    The branch cuts emerge because  $\chi (q, \Omega'_m)$ has a non-analytic, $|\Omega'_m|$
      dependence on
      the
      frequency. In $\chi (q, \Omega'_m)$, viewed as a function of complex $\Omega'_m$, the branch cut is along the imaginary frequency axis  (for $\omega_m = iz + \delta$,  $|\omega_m| = i z \text{sgn} \delta$),
       and it
  runs along
  both
  positive
   and
   negative parts
    of the imaginary axis.
    A convenient way
     to compute $\Gamma_1$ is then to
     split the integral over $\ve_{\bk'}$ into two integrals over positive and negative $\ve_{\bk'}$.
   For each sign of $\ve_{\bk'}$,
   we
     close the integration contour in that
     half-plane which does not contain poles
     and choose the branch cut to be in the same half-plane.
     In this way, only the integral along the branch cut contributes to the final result.

   A closer look at the integral over the  branch cut shows that it is  controlled by energy scales that are much larger than $\Omega_m$ and
   therefore
   can be evaluated
   at $\Omega_m=0$.
    One  such scale
    is $\omega_{{\rm FL}}$, defined in Eq.~(\ref{ofl}),
   and another one is $\omega_0$, defined in Eq.~(\ref{om0}). Note that  $\omega_0\sim\omega_{{\rm FL}} \lambda^3\gg\ofl$.
  The contributions from $\omega_m \sim \omega_{{\rm FL}}$ and from $\omega_m \sim \omega_0$ can be computed
     independently from each other and yield
    $\Gamma_1 = \Gamma_{1,\omega_0} + \Gamma_{1,\ofl}$.
     After some involved algebra, we find
    \bea
   && \Gamma_{1,\omega_0} = \frac{2}{3\pi} \int_0^\infty \frac{dx}{x^{2/3} + x^{4/3} + \sqrt{3} x} =
    \frac{2}{3}, \nonumber \\
    &&
     \Gamma_{1,\omega_{{\rm FL}}} =
     C - \frac{1}{\lambda +1},
    \label{5}
    \eea
 where $C$ is independent
 of
  $\lambda$. This term
  was obtained numerically because
  an analytic
  form of $\Sigma (\omega_m)$ at
  finite $M$ and arbitrary $\omega_m$ is not known.
  However, numerical evaluation of $C$ is straightforward, and
  we found that
  $C=1/3$
  to
  high numerical accuracy.
   The two contributions to $\Gamma_1$ then add up to
 \beq
 \Gamma_1 = \frac{\lambda}{1+ \lambda}.
 \label{ac_3}
 \eeq
 This is  the same result as before,
 but now $\Gamma_1$ comes from  energies which are much higher than $\Omega_m$.

 Taken at
  face value, Eq.~(\ref{5})
  implies
   that $\Gamma_1$ comes
 partially
  from
     $\Omega'_m \sim \omega_{{\rm FL}}$ and
     partially
      from $\Omega'_m \sim \omega_0$. On a more closer look,
      however,
      we found that there is a peculiar cancellation between
       $\Gamma_{1,\omega_0}$ and a portion of $\Gamma_{1,\omega_{{\rm FL}}}$. Namely, $\Gamma_{1,\omega_{{\rm FL}}}$ can be split into two contributions
        -- one is obtained by approximating the fermionic self-energy by  $\Sigma (\Omega'_m) = i\lambda  \Omega'_m$, and another is obtained by subtracting $i \lambda \Omega'_m$ from $\Sigma (\Omega'_m)$.
       In both terms, typical internal frequencies are of order $\ofl$, and the two expressions are of the same order because at $\Omega'_m \sim \ofl$, $\Sigma (\Omega'_m)$ differs from  $i \lambda \Omega'_m$ by terms of comparable magnitude.
     Evaluating the two  parts of $\Gamma_{1,\omega_0}$ separately, we find that the first one gives $\frac{\lambda}{1+ \lambda}$,
     while
     the second gives  $- \frac{2}{3}$ and cancels out $\Gamma_{1,\omega_0}$.  This indicates that the contribution to $\Gamma_{1}$ can be viewed as coming entirely from the range $\Omega'_m \sim \omega_{{\rm FL}}$.
     Still,
     what
     is essential
     for
      our purposes
      is
      that characteristic $\Omega'_m \sim \omega_{{\rm FL}}$ in the vertex correction diagrams is larger than  characteristic $\Omega'_m \sim \Omega_m$ in the internal parts of diagrams {\em c} and {\em d} in Fig.~\ref{fig:side}.  We re-iterate that this separation of scales only holds if we integrate over fermionic frequency first and then over fermionic dispersion.

     The difference between characteristic $\Omega'_m$ and $\ve_\bk$ in the vertex correction diagram in the FL regime also
     holds
     if one calculates vertex corrections in
      the opposite, static  limit, when the external momentum $Q$ is non-zero, while the external frequency $\Omega_m$ is zero. We discuss this issue in
      Appendix \ref{sec:appA}.

\subsubsection{
Final  result for the conductivity in the Fermi-liquid regime}
\label{sec:full}

The separation between characteristic energies in
those parts of
the
current-current correlator, which determine $\gamma_{\text{tr}}$, and
those, which determine vertex corrections, justifies the decomposition of the full correlator, given by diagrams {\em a} and  {\em b}  in Fig.~\ref{fig:side}, into the sum of diagrams {\em c} and {\em d}.
 The internal parts and side vertices in  diagrams {\em c} and {\em d} are computed 
 independently of
  each other.
The
final  result for the  conductivity
 in
 the
 FL regime
 is then
 rigorously established to be
 \beq
 \sigma' (\Omega) \propto \frac{(
 \Gamma Z)^2}{M^2} \sim \frac{1}{M^2} \sim \frac{1}{\omega^{2/3}_{\text{FL}}}.
 \label{8}
 \eeq

  \subsection{Nematic quantum critical point}
 \label{sec:nem_qcp}
At the QCP, $\omega_{{\rm FL}} =0$, and the separation of energy scales does not hold.  Still, Eq.~(\ref{8}) does allow
one to determine $\sigma (\Omega)$ even at
the QCP
 under an additional assumption
that $\omega_{{\rm FL}}$ is the only 
energy
scale near a nematic QCP.
 This assumption is consistent with perturbative calculations.
Combining
this assumption with Eq.~(\ref{8}),  we
conjecture
 that
  the conductivity
 behaves  as $\sigma' (\Omega) \propto \omega^{-2/3}_{FL} f(\Omega/\omega_{{\rm FL}})$, with $f(0) =1$.
Another constraint on function $f(x)$ is imposed by the requirement that $\ofl$ should not enter the result in the quantum-critical regime, where $\Omega\gg\ofl$. This is only possible if $f(x)\propto x^{-2/3}$ for $x\to\infty$. This in turn  implies that
$\sigma' (\Omega) \propto \Omega^{-2/3}$ at the QCP,  in agreement with
Refs.~\onlinecite{kim:1994} and
\onlinecite{eberlein:2016}.

 The scaling argument can also be cast into the renormalization-group language, if we formally introduce
 a
 lower cutoff in the bosonic momentum along the FS
 at some $q_1 \sim (\gamma \omega_1)^{1/3}$, where $\omega_1$ is larger than $\Omega$ but smaller than $\omega_0$.
 This
 cutoff
 effectively re-introduces the mass into the bosonic propagator at
 the
  QCP.
 As the result,
 the fermionic self-energy $\Sigma (\omega)$ and
 local susceptibility $\chi_L (\Omega)$ become scaling functions of $\omega/\omega_1$ and $\Omega/\omega_1$, and display
 a FL behavior at $\omega, \Omega \ll \omega_1$.  Accordingly, the conductivity scales as $\sigma'(\Omega) \propto \omega^{-2/3}_1$.  One can then make $\omega_1$ progressively smaller and get progressively larger conductivity.  The scaling $\sigma'(\Omega) \propto \omega^{-2/3}_1$
  holds as long as $\omega_1 \gg \Omega$.
  At $\Omega\lesssim \omega_1$, scaling with $\omega_1$ is replaced by that with $\Omega$, which
  yields again $\sigma'(\Omega) \propto \Omega^{-2/3}$.

 \section{Spin-density-wave\\ quantum critical point}
 \label{sec:SDW}

\begin{figure}[h]
\vspace{-0.1in}
\includegraphics[width=1.0 \linewidth]{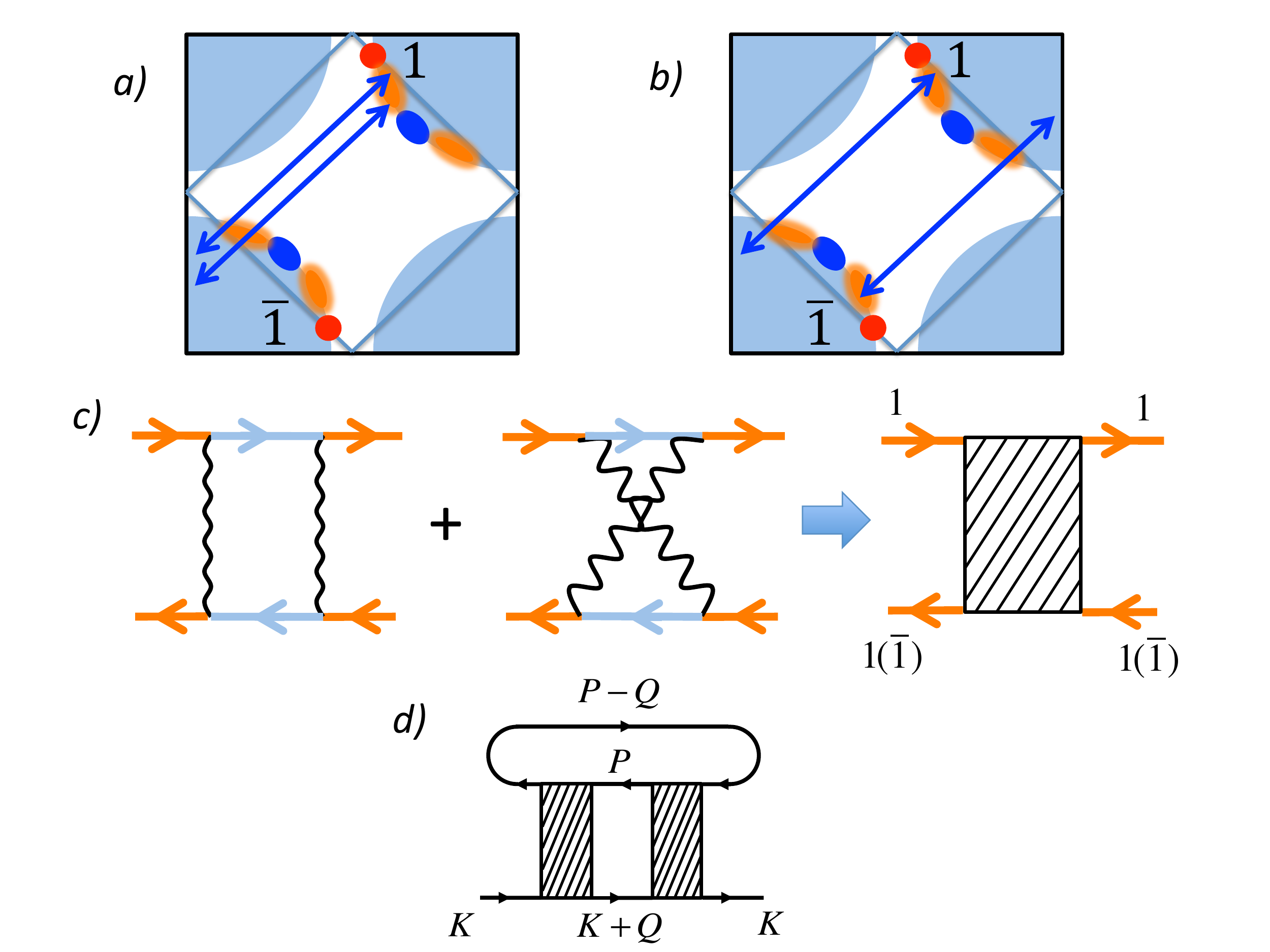}
\vspace{-0.2in}
\caption{{\em a} and {\em b}: A Fermi surface of a 2D metal near a SDW QCP. Red circles: hot spots; blue areas: cold regions; orange areas: lukewarm regions. Arrows indicate a composite scattering process which involves lukewarm fermions either from the same ({\em a}) or diametrically opposite ({\em b}) regions. {\em c}: Composite scattering vertex. The initial states on the bottom can belong either to lukewarm regions $1$ or $\bar 1$. {\em d}: Two-loop composite self-energy.\label{fig:comp_SDW}}
\end{figure}

 The correlation function of antiferromagnetic fluctuations near a SDW QCP,
 \beq\label{chi_q}
 \chi (\bq, \Omega_m) = \frac{\chi_0}{(\bq- \bq_{\pi})^2 + M^2 + \gamma |\Omega_m|},
 \eeq
  is peaked at the nesting momentum $\bq_\pi$ which connects hot spots on the Fermi surface. For a 2D square lattice, $\bq=(\pi,\pi)$ (the lattice constant is set to unity). Two out of eight hot spots are shown by red circles in Fig.~{\ref{fig:comp_SDW}}, panels {\em a} and {\em b}.  In what follows, we will consider the contributions to the optical conductivity both from ``hot fermions", located near the hot spots, and from ``lukewarm fermions'',\cite{hartnoll:2011,chubukov:2014} occupying the regions between the hot spots and cold parts of the FS (the cold and lukewarm regions are depicted as blue and orange areas, correspondingly, in Fig.~{\ref{fig:comp_SDW}} {\em a} and {\em b}).

  \subsection{Conductivity of hot fermions}
  \label{sec:hot}
The main interaction mechanism for hot fermions is SDW scattering by momentum $\bq_\pi$, mediated by the effective interaction in Eq.~(\ref{chi_q}).
To one-loop order,
this interaction leads
 to  a
 singular behavior of the self-energy
 at  the hot spots
 (${\bf k} = \bk_{\text{hs}}$),
  where  $\Sigma (\bk_{\text{hs}}, \omega) \propto \omega^{1/2}$.
 Away from
 the
  hot spots, this singular behavior holds in a range of $
 |{\bf k} - \bk_{\text{hs}}|$
 whose width by itself scales as  $\sqrt{\omega}$.  This additional factor
 of $\sqrt{\omega}$
 can be incorporated into the  transport scattering rate and, beyond that, does not affect our consideration.
 For scattering peaked at the
 nesting
  momentum,
 the velocities at two hot spots connected by $\bq_\pi$, $\bv_\bk$ and $\bv_{\bk+
 \bq_\pi} $,
 have equal magnitudes but generally differ in direction.
 On one hand, this implies that
the factor
 $(\bv_\bk - \bv_{\bk+
 \bq_\pi})^2$
 in
 the transport scattering rate does not introduce additional smallness, i.e., $
 \gamma
 _{\text{tr}}$ and FS-averaged $\Sigma'' (\omega)$ are of the same order.  On the other hand,
 renormalizations of the current vertices at the initial  and final  points of a SDW scattering process (${\bf k}$ and ${\bf k} + \bq_\pi$, correspondingly)
 are mixed in the perturbation theory.
  \begin{figure}[htp]
\vspace{-0.3in}
\includegraphics[width=1.0 \linewidth]{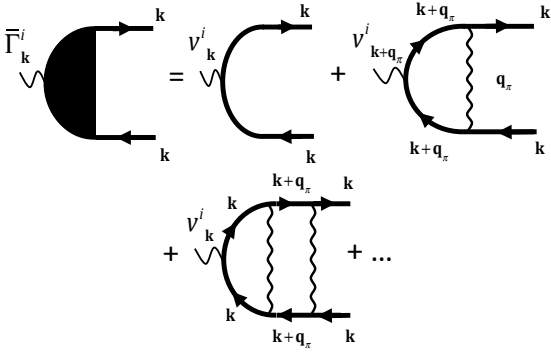}
\vspace{-0.6in}
\caption{Diagrammatic series for the current vertex near a SDW QCP. $\bk$ is chosen at a hot spot
and $\bk+\bq_\pi$ is at another hot spot, connected to the first one by the
nesting
 momentum, $\bq_\pi=(\pi,\pi)$. The superscript $i$ denotes the $i^{\text{th}}$ Cartesian component of the corresponding vector.
The series for the vertex at $\bk+\bq_\pi$ is obtained from that shown in the figure by relabeling $\bk\leftrightarrow\bk+\bq_\pi$. \label{fig:vertex_SDW}}
\end{figure}

 To get an insight
 into
 this mixing,
 we
 consider again the FL regime,  where the self-energy near a hot spot is described by $\Sigma
(\omega_m) = i \lambda \omega_m$, and
vertex renormalization
is described by a geometric series of ladder diagrams.
The series for the $i^{\text{th}}$
 Cartesian
component of $\boldsymbol{\bar\Gamma}_{\bk}$
is
 shown in Fig.~\ref{fig:vertex_SDW}; the series for $\boldsymbol{\bar\Gamma}_{\bk+\bq_\pi}$
 is
  obtained  by relabeling $\bk\leftrightarrow\bk+\bq_\pi$.
Performing the same calculations as for the nematic case, we obtain
 \bea
 && {\bar \Gamma}^i_\bk = \frac{v^i_\bk + \frac{\lambda}{1 + \lambda} v^i_{\bk + \bq_\pi}}{1 - \left(\frac{\lambda}{1 + \lambda}\right)^2}, \nonumber \\
&& {\bar \Gamma}^i_{\bk + \bq_\pi} = \frac{v^i_{\bk+ \bq_\pi} + \frac{\lambda}{1 + \lambda} v^i_{\bk}}{1 - \left(\frac{\lambda}{1 + \lambda}\right)^2}.
 \label{9}
 \eea
 Each of the vertices in Eq.~(\ref{9}) diverges at criticality, where $\lambda\to\infty$. However, one can
 readily
  verify that for
  the
  SDW case the sum of diagrams
 {\em c} and {\em d} in Fig.~\ref{fig:side} is equal to diagram {\em c} with side vertices
 ${\bar \Gamma}^i_\bk - {\bar \Gamma}^i_{\bk + \bq_\pi}$.
  From Eq.~(\ref{9}) we see that this difference is finite at $\lambda\to\infty$:
 \beq
 {\bar \Gamma}^i_\bk - {\bar \Gamma}^i_{\bk + \bq_\pi} = \frac{v^i_\bk - v^i_{\bk + \bq_\pi}}{1 + \frac{\lambda}{1 + \lambda}} \approx \frac{1}{2} \left(v^i_\bk - v^i_{\bk + \bq_\pi}\right).
 \label{10}
 \eeq
 Therefore
 the
 effective
 current vertex,
 which appears in the expression for the conductivity,
 does not undergo singular renormalization.
 As a result,
 the fermionic $Z$-factor does not cancel out from the conductivity,
 and we have
 $\sigma'(\Omega) \propto Z^2\gamma_{\text{tr}}/\Omega^2
 $.
 [We remind that in our local theory there is a one-to-one correspondence between the  $Z$-factor and the renormalized mass: $Z=m_b/m^*$.
 It is expected that in a more general case, when this relation does not hold, the $Z$-factor is to be replaced by $m_b/m^*$.]
 In
 the
   FL regime,
 $Z
 \propto
  M$ and
 $\Sigma''(\Omega)\propto \Omega^2/M^3$.
 The
 role
 $\gamma_{\text{tr}}$
 is played by the FS average of $\Sigma''$,
 which differs from $\Sigma''$ by the angular width of the hot spot.
 This width by itself scales as $M$; thus $\gamma_{\text{tr}} \propto \Omega^2/M^2$.
 Collecting all the factors together, we find that
$\sigma'(\Omega)$ tends to
an
$M$-independent value
 at $\Omega \to 0$.
At the
QCP, $Z \propto \sqrt{\Omega}$, $\gamma_{\text{tr}} \propto \sqrt{\Omega}\times \sqrt{\Omega} = \Omega$, and $\sigma' (\Omega)$ again
remains
constant at
$\Omega\to 0$. \footnote{A frequency-independent conductivity in $D=2$ would be consistent  with the hypothesis of hyperscaling,\cite{patel:2015} according to which $\sigma \propto \Omega^{\frac{D-2}{z}}$, where $z$ is the dynamical critical exponent. This agreement  may, however, be accidental  as it does not hold, within the same reasoning,  beyond $D=2$.}

The results presented above imply that the extended Drude formula [Eq.~(\ref{ch_1})] correctly describes the  hot-fermion conductivity.
  Higher-order self-energy and vertex corrections do contain
additional factors of
$\ln{\Omega}$, and a series of such terms may give rise  to a singular behavior of the hot-fermion conductivity.  Still, Eq.~(\ref{ch_1}) is expected to be valid except, possibly, very
 frequencies.

\subsection{Conductivity of lukewarm fermions}
\label{sec:warm}

 Previous studies\cite{hartnoll:2011,chubukov:2014,maslov:2017} found that the hot-spot contribution to
 the
 conductivity is not the dominant one at low frequencies,
  when there is a clear distinction between hot and cold regions
  of
   the Fermi surface (at high enough frequencies, the full Fermi surface becomes ``hot"~\cite{abanov:2003}).  The dominant contribution to
   the
    optical conductivity actually comes from lukewarm regions, located in between hot and cold regions on the Fermi surface.
Fermions in the lukewarm regions (orange areas in Fig.~\ref{fig:comp_SDW}, {\em a} and {\em b}) form a FL state even if the system is right at the SDW criticality. However, this is a strongly renormalized FL with a $Z$-factor which varies
 from zero at the hot spot to $Z\approx 1$ in the cold region.
In the bulk of the lukewarm region, the $Z$-factor scales linearly with the distance along the FS measured from the nearest hot spot: $Z\sim k_{||} v_F/\bar g\ll 1$. The most relevant interaction process for lukewarm fermions is composite scattering,\cite{hartnoll:2011} which consists of
two consequent events of scattering by $\bq_\pi$. Because a lukewarm fermion is not at the hot spot,
the first scattering event by $\bq_\pi$ takes it to
an off-shell state away from the FS, and the second event brings it back to near where it started. In principle, fermions of all the eight hot spots
can be involved in composite scattering,  but the the corresponding two-loop self-energy (Fig.~\ref{fig:comp_SDW}{\em d}) is logarithmically enhanced in  two cases:
if the lukewarm fermions
belong to same region (``forward scattering'', shown in Fig.~\ref{fig:comp_SDW}{\em a}) or diametrically opposite regions (``$2k_F$-scattering'', shown in Fig.~\ref{fig:comp_SDW}{\em b}). The corresponding scattering
vertices are shown in Fig.~\ref{fig:comp_SDW}{\em c}. For lukewarm fermions at distances $p_{||}$ and $k_{||}$ from the corresponding hot spot(s),
the composite vertex with momentum transfer $q$ and frequency transfer $\omega$ is of order $\Gamma_{\text{c}}\sim (\bar g/k_{||} p_{||}) \ln\left(\Lambda/\max\{\omega,v_Fq\}\right)$.

The most singular contribution to the optical conductivity occurs at two-loop order in composite scattering. Depending on the energy the system of lukewarm
fermions is probed at, it behaves either as a 1D or 2D system. For the optical conductivity, the energy scale separating the two regimes is $\Omega_{12}\sim \bar g^2/E_F$.

For $\Omega>\Omega_{12}$, the energy cost of displacing a lukewarm fermion  tangentially to the FS is small, which means that the curvature of the Fermi surface can be neglected, and we are in the 1D regime. The corresponding self-energy exhibits a linear scaling with frequency $\Sigma''\propto\Omega$, which is characteristic for 1D.\cite{maslov:2004}
The main contribution to $\sigma'(\Omega)$ in this regime
comes from the boundary between the lukewarm and cold regions of the FS, where $Z\sim 1$ (Refs.~\onlinecite{hartnoll:2011,chubukov:2014,maslov:2017}). In this case, Eq.~(\ref{ch_1_1}) with $\gamma_{\text{tr}}\sim\Sigma''\propto\Omega$ predicts that
\bea
\sigma'(\Omega)\propto 1/\Omega.\label{1D}
\eea

For $\Omega<\Omega_{12}$, the FS curvature cannot be neglected, and we are in the 2D regime.
The main contribution to $\sigma'(\Omega)$ comes from
the
 region of
 $k_{||}
  \propto \Omega^{1/3}$ (Refs.~\onlinecite{hartnoll:2011,chubukov:2014,maslov:2017}),  where the $Z$-factor is small: $Z\propto
  k_{||}
  \propto \Omega^{1/3}\ll 1$.
  Therefore, the question whether renormalization of the $Z$-factor affects $\sigma'(\Omega)$ is again relevant.

\begin{figure}[htp]
\vspace{-0.15in}
\includegraphics[width=1.0 \linewidth]{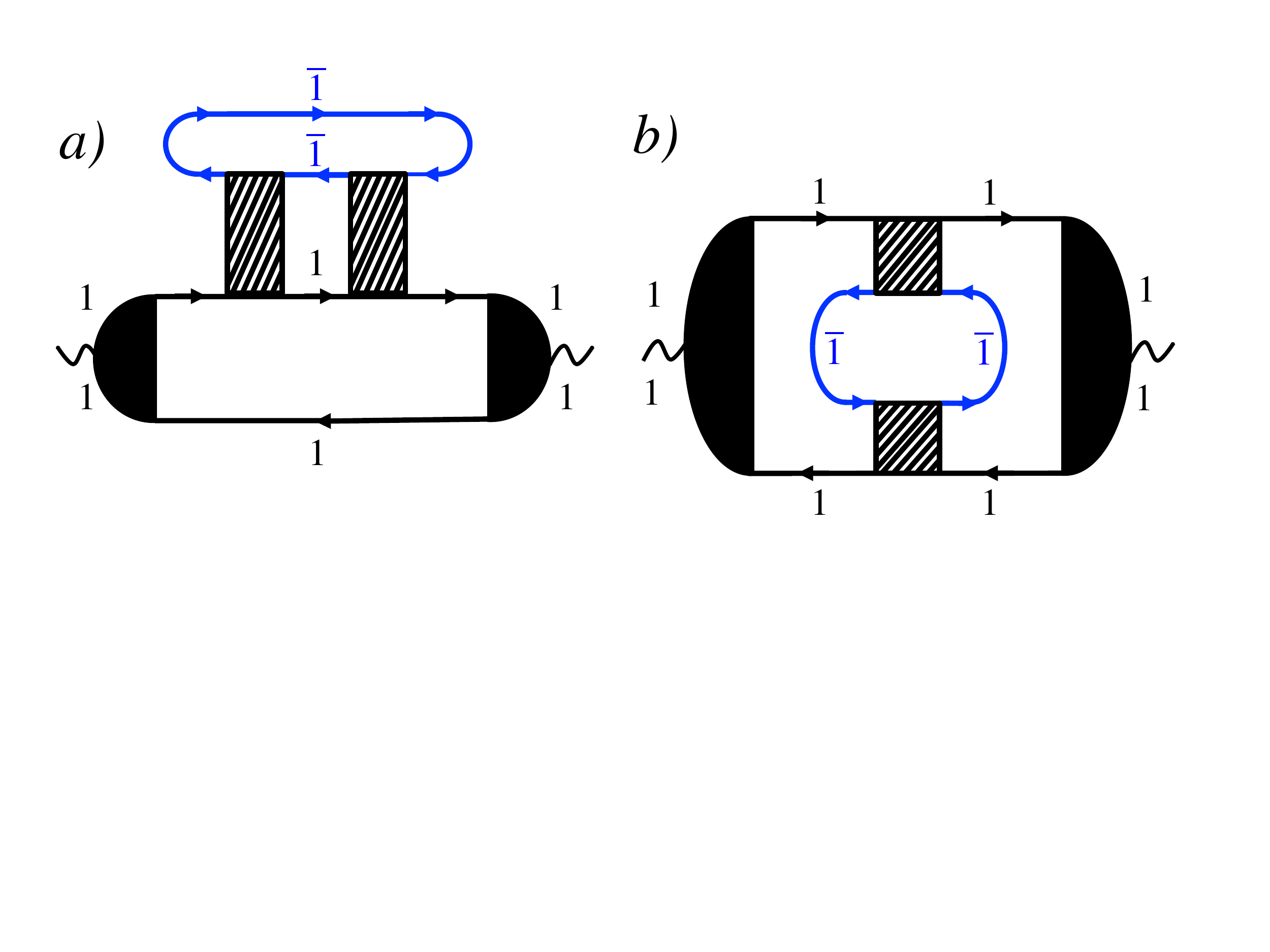}
\vspace{-1.2in}
\caption{Maki-Thompson diagrams for the conductivity of lukewarm fermions (the mirror image of diagram {\em b} is not shown.) Hatched boxes represent composite scattering vertices shown in Fig.~\ref{fig:comp_SDW}{\em c}. Current vertices are renormalized by $\bq_\pi$ scattering as shown in Fig.~\ref{fig:vertex_SDW}. Labels $1$ and $\bar 1$ correspond to
the lukewarm regions in Fig.~\ref{fig:comp_SDW}, {\em a} and {\em b}. \label{fig:comp_cond}}
\end{figure}

The two-loop self-energy in the 2D regime is of the FL type
 $\Sigma''(\Omega)\propto (\Omega^2/k_{||}^4)
 \ln^3
( \Omega_{12}/\Omega)$ (the factor of $1/k_{||}^4$ comes from the product of two composite vertices in Fig.~\ref{fig:comp_SDW}{\em d}).  The corresponding
Maki-Thompson diagrams for the conductivity are shown in Fig.~\ref{fig:comp_cond}.
(As before, we neglect the Aslamazov-Larkin diagrams which would
only modify the result by a factor of order one because our system is on a lattice.)
 The current vertices in these diagrams are formed by one-loop $\bq_\pi$ scattering, which is still the main process leading to renormalization of the $Z$-factor. However, although the composite vertices (hatched blocks) 
are 
constructed from two $\bq_\pi$ scattering processes, they effectively scatter fermions only by small angles. In Fig.~\ref{fig:comp_cond}, we depicted a particular $2k_F$ composite scattering processes, in which the two incoming fermions  belong to diametrically opposite lukewarm regions ($1$ and $\bar 1$).
    The current vertices in both diagrams {\em a } and {\em b} belong to the same lukewarm region ($1$). In diagram {\em a}, the left and right current vertices are evaluated at the same momentum.  In diagram {\em b}, the momenta in the left and right current vertices  differ by a small momentum transfer through the composite vertex. In this sense, the situation is now similar to the nematic QCP but partial cancellation between
diagrams {\em a} and {\em b} affects only the logarithmic factors in the self-energy.~\cite{chubukov:2014,maslov:2017}
  As a result, $\gamma_{\text{tr}}(\Omega)\sim \Sigma''(\Omega)/\ln^3(\Omega_{12}
/\Omega)
  \propto \Omega^2/k_{||}^4$.
  If renormalization of the current vertices is neglected,
   the conductivity is obtained from Eq.~(\ref{ch_1_1}) by replacing $Z$ and $\gamma_{\text{tr}}(\Omega)$ by their values at given $k_{||}$ and averaging over $k_{||}$.
The lower limit of the momentum integration is $k
_{||}
\sim \Omega^{1/3}$,
while
 the upper limit 
can be set to infinity due to a rapid convergence of the integral.
Then we would obtain
   $\sigma'(\Omega)\propto \int^\infty_{
   \Omega^{1/3}
   } dk_{||} Z^2\gamma_{\text{tr}}(\Omega)/\Omega^2
    \propto 1/\Omega^{1/3}$. This is the result reported in
   Refs.~\onlinecite{hartnoll:2011,
  chubukov:2014,maslov:2017}.

  We now follow the analysis of a nematic QCP  and
  take renormalization of the current vertices into account.
   Each of the current vertices diverges at criticality as specified by Eq.~(\ref{9}), i.e., $\bar\Gamma^i_{\bk}\propto \lambda\sim 1/Z\propto 1/k_{||}$.
 Consequently,
 the result for the conductivity is changed to
\bea
\sigma'(\Omega)
\propto
 \frac{1}{\Omega^2}\int^\infty_{\Omega^{1/3}} dk_{||}\left(\bar\Gamma^{i}_{\bk}Z\right)^2\gamma_{\text{tr}}(\Omega)
\propto
 \frac{1}{\Omega}.\label{2D}
\eea
The
key part of the
this result
is a cancellation between the $Z$-factor and $\bar\Gamma^{i}_{\bk}$ which, as for the nematic case, leads to a breakdown of the extended Drude formula [Eq.~(\ref{ch_1})].
As the consequence,  the
$1/\Omega$ scaling of $\sigma'(\Omega)$
 extends from the
 1D regime [Eq.~(\ref{1D})] down to the lowest frequencies.

Comparing the hot-spot and lukewarm contributions to the conductivity, we see that the latter
 is much larger, both in
 the
 2D and
  1D regimes.
 Eventually, the full Fermi surface becomes hot,~\cite{abanov:2003} but this happens only at high enough frequencies.
 Therefore, $\sigma'(\Omega)$ at a SDW QCP  scales as $1/\Omega$.
    We note in passing that such
    scaling is consistent with the marginal-FL phenomenology \cite{varma:1989, *varma:2002} and observed scaling of $\sigma'(\Omega)$ in the high-$T_c$ cuprates. \cite{basov:1996, *puchkov:1996, *norman:2006}

 \section{Conclusions}
 The question of how
 renormalization of the electron effective mass
 affects the conductivity  has a long history, which goes back  to the seminal papers  by Langer  on the residual resistivity of a FL\cite{langer:1960} and
 by Langreth and Kadanoff on polaronic transport. \cite{langreth:1964}
 Nowadays, this question has acquired particular importance
 in the context of correlated
electron systems near
quantum phase transitions,
where the renormalized mass
is expected to depend on the temperature or  frequency, thus potentially affecting the
corresponding dependences of the conductivity.  A phenomenological way to account for these extra dependences is via the ``extended Drude formula''\cite{basov:2011} of the type given by Eq.~(\ref{ch_1}), which contains the renormalized (and thus $\Omega$- and $T$-dependent) mass.

In this paper, we studied the optical conductivity  within two specific models of quantum criticality of the nematic and spin-density-wave types. In both cases, the critical theory is local, and the
effective mass is same as the  inverse quasiparticle $Z$ factor.
We found that in the nematic case the effect of mass renormalization is canceled by renormalization of current vertices. This is consistent with earlier works.~\cite{kim:1994,eberlein:2016} The resultant conductivity, $\sigma'(\Omega)\propto\Omega^{-2/3}$ is consistent with the Drude formula which contains
a bare rather than
 renormalized mass.
The spin-density-wave case happens to be more subtle. There are two contributions to the conductivity from two qualitatively different regions of the Fermi surface: hot spots, connected by the nesting vector, and lukewarm regions, occupying the space in between hot and cold parts of the Fermi surface.
We found no
cancelation between the mass  and current vertex for the hot-fermion contribution.
In this situation,  the correct result for the conductivity is reproduced by the Drude formula with the renormalized mass. On the other hand, this cancellation
does occur for the lukewarm-fermion contribution,
 which is the
 dominant
 one.
The resultant conductivity at a SDW QCP is then $\sigma'(\Omega)\propto 1/\Omega$.

In view of these results, we believe that the short answer to the question "bare
vs renormalized mass"  is "it depends on
the situation considered".
 For example, the {\em dc} conductivity of electrons coupled to optical phonons contains the bare mass in the adiabatic regime, when the electron energy
 is higher than the phonon one,  and the renormalized mass in the anti-adiabatic regime,\cite{langreth:1964} when the electron energy
 is lower than the phonon one. The
 three
 cases that we considered  here
provide
 three
 more examples which demonstrate the absence of the universal answer to
 the question about the type of
 the effective mass
  entering
   the conductivity.  Indeed, whether the conductivity of a quantum-critical system contains the bare or renormalized mass turns out to depend
   not only
    on the type of criticality ($q=0$ vs finite $q$ QCP),
    but also on particular scattering processes  considered for a given type.
    Overall,
    one
     implication of our study is that  the extended Drude formula need to be treated with a great caution.

     One more reason  for exercising caution
     is that Eq.~(\ref{ch_1}) is not the only form of the extended Drude formula.
      Another version of this formula can be derived from the kinetic equation for a FL:\cite{nozieres:1966,physkin}
    \bea
    \sigma'(\Omega)=\frac{\Omega_p^2}{4\pi}\frac{m_b}{m^*}\frac{\tilde\gamma_{\text{tr}}(\Omega)}{\left(\Omega\frac{m_b}{m^*}\right)^2+\tilde\gamma^2_{\text{tr}}(\Omega)}.\label{DFL}
    \eea
    In this version, the frequency in the denominator is {\em divided} by the renormalized  mass, which is opposite to what Eqs.~(\ref{ch_1}) says.  \footnote{In the original derivation,\cite{nozieres:1966} $\gamma_{\text{tr}}$ is due to impurity scattering but we adopt the same form, assuming momentarily that it holds also for other types of scattering.} Note, however, the transport scattering rate in Eq.~(\ref{DFL}) is introduced phenomenologically and cannot be {\em a priori} associated with the fermionic self-energy, even if a transport correction is accounted for. As pointed out in Ref.~\onlinecite{kim:1994}, Eq.~(\ref{DFL}) can be made consistent with the result $\sigma'(\Omega)\propto \Omega^{-2/3}$ for a nematic QCP by redefining the transport scattering rate
    as $\gamma^*_{\text{tr}}=\tilde\gamma_{\text{tr}}(m^*/m_b)$ and assuming that it is $\gamma^*_{\text{tr}}$ rather than
   $\tilde\gamma_{\text{tr}}$  that scales as $\Omega^{4/3}$ at criticality. Such a redefinition changes Eq.~(\ref{DFL}) to
   $
    \sigma'(\Omega)=\Omega_p^2\gamma^*_{\text{tr}}/\left[4\pi\left(\Omega^2+\gamma^{*2}_{\text{tr}}\right)\right]
   $
        which does not contain the renormalized mass.

\acknowledgements
We thank
S. Hartnoll, Y.B. Kim and  S. Sachdev for fruitful discussions.
This work was supported by the NSF DMR-1523036 (A.V.C.).
We acknowledge the hospitality of the Kavli Institute for Theoretical Physics, which is supported by the NSF via Grant No. NSF PHY11-25915.

\appendix

\section{Vertex corrections at
finite external momentum and zero external frequency}
\label{sec:appA}
For completeness,
we analyze in this Appendix
vertex renormalization
  in the situation when the
the incoming frequency is set to zero
 while keeping
 the
 external momentum $Q$  finite but small ($\Omega/Q\to 0$).
 In this limit, the Ward identity relates the current vertex to the fermionic self-energy via\cite{engelsberg:1963}
 \bea
 \boldsymbol{\bar\Gamma}^Q=1-\nabla_{\bk}\Sigma(\bk,\omega=0).\label{dm_1}
 \eea
 In an isotropic system, one can write $\Sigma(\bk,\omega)=\Sigma(\ve_\bk,\omega)$ and $\boldsymbol{\bar\Gamma}^Q=\bv_\bk \Lambda$, where $\Lambda$ depends only on the magnitude of $\bk$. Furthermore, if we restrict ourselves to nematic criticality, where typical momentum transfers are small, $\Lambda$ coincides with the density vertex in the $\Omega/Q\to 0$ limit, which we will denote by $\Gamma^Q$.
 Then Eq.~(\ref{dm_1}) is reduced to
\beq
\Gamma^Q = 1
- \frac{\partial \Sigma (\ve_\bk, \omega_m)}{\partial{\ve_\bk}}\Big\vert_{\omega
=0,
\ve_\bk \to 0}.
\label{ac_1}
 \eeq

 \subsection{Eliashberg approximation}
 In the
 main text, we used an approximate scheme to compute the self-energy $ \Sigma (\bk, \omega)
 $ for $k = k_F$.
  Namely, we factorized
 the 2D internal momentum ${\bf k}_F + {\bf q}$ into components tangential and normal to the FS
 and kept
 the dependence on $q_{\perp}$ only in
 the Green's function, in which $\ve_{\bk_F+\bq} = v_F q_{\perp}$, and neglected
$q_{\perp}$ in the bosonic propagator, leaving it as a function of  $q_{\parallel}$ only.
   The reasoning was that characteristic  $q_{\perp}$ are much smaller than characteristic  $q_{\parallel}$ both at a QCP in the FL region near a QCP. This approximation is similar to the Eliashberg
   approximation used analysis of electron-phonon interaction.
   Within
   this approximation,
     $\Sigma (\ve_\bk, \omega_m =0)$ is independent of $\ve_\bk$.
     According to Eq.~(\ref{ac_1}), the
     vertex
     is then not renormalized,
     i.e., $\Gamma^Q=1$.

  The absence of
  renormalization of $\Gamma^Q$ becomes immediately evident if we evaluate the building block of the ladder
  series by integrating over the fermionic dispersion first. The building block of the series
  is similar to that in Eq.~(\ref{gamma1}),
  except for now the external frequency is zero and the external momentum is finite. The first-order correction to the vertex is
  given by
    \bea
\Gamma^Q_1 &=& g^2
\int  \frac{d\Omega'_m}{2\pi} \int  \frac{d^2q}{(2\pi)^2}
G(\bk_F + \bq, \Omega'_m) G(\bk_F + \bq + \bQ, \Omega'_m)\nn\\
&&\times \chi (q,\Omega'_m).\label{gamma1q}
  \eea
  The
  Green's functions are the full ones: $G(\bk, \Omega'_m) = \left[i \Omega'_m + \Sigma (\Omega'_m) - \ve_\bk \right]^{-1}$. The
  fermionic dispersions are $\ve_{\bk_F + \bq} = v_F q_{\perp}$ and $\ve_{\bk_F + \bq + \bQ} = v_F \left(q_{\perp} + Q_{\perp}\right)$, and we approximate $\chi (q,\Omega'_m)$ by $\chi (q_\parallel,\Omega'_m)$.
  The
  poles in the Green's function are located in the same half-plane of complex $q_{\perp}$, hence the integral over $q_\perp$ vanishes.

  The same result can be obtained by integrating
  over $\Omega'_m$ first. Now $\Gamma^Q_1$ has two contributions. One comes from the poles in the
  Green's functions and another from the branch cut in the bosonic propagator.  The pole contribution is non-zero
  if $q_{\perp}$ and $q_{\perp} + Q_{\perp}$ have different signs, i.e.,
  if
  $-Q_\perp<q_{\perp}<0$
  (for $Q_{\perp}>0$). When the limit of $Q \to 0$ is taken,
  the width of this interval shrinks to zero, hence the poles are located at vanishingly small $\Omega'_m$. For such $\Omega'_m$, the self-energy can be approximated by $i \lambda \Omega'_m$, i.e.,
  the
   Green's function  can be approximated by   $G(\bk, \Omega'_m) = \left[i \Omega'_m (1 + \lambda) - \epsilon_k \right]^{-1}$. At the same time, $\chi (q_\parallel,\Omega'_m)$ can be approximated by $\chi (q_\parallel,0)$.
   Evaluating the integral over $\Omega'_m$ and then two independent integrals over $q_{\perp}$ and $q_{\parallel}$, we find
   \beq
   \Gamma^Q_{1,\text{poles}} = - \frac{\lambda}{1+ \lambda}.
   \label{ac_2}
   \eeq
   The contribution from the branch cut does not
   depend on the order of limits $\Omega_m\to 0$ and
   $Q\to 0$,
   and is given by Eq.~(\ref{ac_3}):
    \beq
   \Gamma^Q_{1,\text{br-cut}} = \frac{\lambda}{1+ \lambda}
   \label{ac_4}
   \eeq
  Adding Eqs.~(\ref{ac_2}) and (\ref{ac_4}),
  we find that
  $\Gamma^Q_1$ vanishes, as we also found by integrating over the dispersion first.    Higher-order vertex corrections can be computed in the same way, and also vanish.  As a result, within Eliashberg approximation, the ladder series reproduce the Ward identity $\Gamma^Q =1$.

  A comment is in order here. At
   first glance, the vanishing of the sum of $\Gamma^Q_{1,\text{poles}}$ and $\Gamma^Q_{1,\text {br-cut}}$ implies that vertex corrections are not needed
   for
   a diagrammatic derivation of
the FL results for  the uniform static charge and spin susceptibilities
 $\chi_{c,s}
= 2N_F (m^*/m)/(1 + F_{0}^{c,s}) = 2N_F (1 + F^{c}_1)/(1 + F_{0}^{c,s})$, where $F_{l}^{c,s}$ are Landau parameters in the charge ($c$) and spin ($s$) channels.
  Diagrammatically, $\chi_{c,s}$
   are
   given
    by a fully renormalized polarization bubble with zero external frequency and small but finite $Q$, and the vertices in such a bubble seem to be $\Gamma^Q$.
    However,
     in a diagrammatic calculation one explores the separation of scales and absorbs all contributions coming from finite internal frequencies and momenta into Landau parameters,
     which play a role of irreducible vertices in diagram Fig.~\ref{fig:side}{\em d} and its extensions to higher orders
     (see Refs.~\onlinecite{
     finkelshtein:2010,chubukov:2014b}).
     These parameters are then used as inputs for the computations of the contribution coming from infinitesimally small internal momenta and frequencies.   Within this approach, $\Gamma^Q_{
     \text{br-cut}}$
     contributes to the Landau parameters, while
  $ \Gamma^Q_{1,\text{poles}}$ contributes the to middle sections of the diagrams, formed by low-energy fermions.
Because $\Gamma^Q_{1,\text{br-cut}}$ is the same as the vertex correction for the opposite case, when $Q=0$ and $\Omega_m$ is small but finite, this contribution is in fact a part of $\Gamma^\Omega_1$, and the series of $\Gamma^Q_{1,\text{br-cut}}$, taken alone, are summed up into $\Gamma^\Omega = 1 + \lambda = Z^{-1}$. The product of two dressed fermion-boson vertices and factors of $Z^2 m^*/m$ from two low-energy
Green's functions
then combine to produce $F^{1}_c$. In the same manner, series of renormalizations of the 4-fermion interaction, all coming from internal energies of order $M$ and therefore insensitive to the interplay between external $\Omega_m$ and $Q$, combine with the $Z^2 m^*/m$  factors to produce  $F^{0}_{c,s}$ in the denominator of the Landau formula for the  uniform susceptibility.

\subsection{Beyond the Eliashberg approximation}

The
 calculation
gets
more involved if one
goes beyond
the Eliashberg approximation and keep $q_{\perp}$ in the bosonic susceptibility.
Then the self-energy $\Sigma (\ve_\bk, \omega_m)$ acquires
a
$\ve_\bk$ term
and its $\omega_m$ term gets a correction:
$\Sigma (\ve_\bk, \omega_m)=i\lambda\omega_m+
A (i\omega - \ve_\bk)$. In contrast to the $\omega_m$ term, whose prefactor diverges
at criticality, the prefactor $A$ is ${\cal O}(1)$ even right at the QCP.
(More precisely,
$A$ acquires a logarithmic dependence on $\ve_\bk$ starting at three loop order,\cite{metlitski:2010b, *metlitski:2010c}
 but we will not dwell into this here.)  Accordingly, when we compute $\Gamma^Q_1$ by integrating over $\ve_\bk$ first, we now find that this term is  non-zero due to the pole in $\chi (q, \Omega'_m)$ viewed as a function of $q_{\perp}$. The vertex correction $\Gamma^Q_1$  is ${\cal O}(1)$, but, unlike
 the
 $\lambda/(1 + \lambda)$ correction to the vertex in the $Q/\Omega_m\to 0$ limit,
   is not close to one.  Accordingly, the series of vertex corrections are expected to sum up into  $\Gamma^Q = {\cal O}(1)$, as the Ward identity implies.  It has not been checked, however, that
  summing only the ladder series of vertex corrections
  reproduces the Ward identity diagrammatically.

 Note in this regard that $\Gamma^Q_1$ can be made small if we formally extend the theory to $N$  fermionic flavors and take the limit $N \gg 1$.  Then $\Gamma^Q_1 = {\cal O}(1/N)$ and one does not need to extend the calculation of $\Gamma^Q$ beyond $\Gamma^Q_1$ to reproduce the Ward identity.
   The large-$N$ expansion is also known to break at three-loop at higher order,\cite{lee:2009} so it does not actually help much from
 from the rigorous point of view.
  For practical purposes, however, multi-loop contributions to $\Sigma (\ve_\bk, \omega_m =0)$ and to $\Gamma^Q$ are rather small numerically, so to a good accuracy one can approximate $\Gamma^Q$ by $1+ \Gamma^Q_1$ and
 $\Sigma(\ve_\bk, 0)$ by the one-loop result $\Sigma (\ve_\bk, 0) =
 -A_1 \ve_\bk$.  To this order,  $\Gamma^Q_1 = A_1$, i.e., the Ward identity
 is reproduced.

  For completeness, we also look at  vertex renormalization at $Q=0$ and $\Omega_m \to 0$ beyond
  the Eliashberg approximation.  Using
 $\Sigma ( \ve_\bk, \omega_m) = i \omega_m \lambda  + A (i \omega_m - \epsilon_k)$ yields
 $\Sigma ( \ve_\bk =0, \omega_m) = i \omega_m \left(\lambda  + A\right)$.
 One-loop vertex renormalization is
 $\Gamma^\Omega_1 = \lambda/(1 + \lambda + A) +A$.
  At the next order, $\Gamma^\Omega_2 = \left[\lambda/(1 + \lambda + A)\right]^2  +A\lambda/(1 + \lambda + A)$.  This suggests that the full series reduce to
  \beq
  \Gamma^\Omega = \frac{1+A}{1- \frac{\lambda}{1+\lambda + A}} = 1 + \lambda + A.
  \label{ac_5}
 \eeq
 This is consistent with the  Ward identity $\Gamma^\Omega = 1 + \partial\Sigma (\ve_\bk=0, \omega_m)/\partial(i \omega_m) = 1 +\lambda + A$.

\bibliography{cond_vertex.bib}

\begin{thebibliography}{54}%
\makeatletter
\providecommand \@ifxundefined [1]{%
 \@ifx{#1\undefined}
}%
\providecommand \@ifnum [1]{%
 \ifnum #1\expandafter \@firstoftwo
 \else \expandafter \@secondoftwo
 \fi
}%
\providecommand \@ifx [1]{%
 \ifx #1\expandafter \@firstoftwo
 \else \expandafter \@secondoftwo
 \fi
}%
\providecommand \natexlab [1]{#1}%
\providecommand \enquote  [1]{``#1''}%
\providecommand \bibnamefont  [1]{#1}%
\providecommand \bibfnamefont [1]{#1}%
\providecommand \citenamefont [1]{#1}%
\providecommand \href@noop [0]{\@secondoftwo}%
\providecommand \href [0]{\begingroup \@sanitize@url \@href}%
\providecommand \@href[1]{\@@startlink{#1}\@@href}%
\providecommand \@@href[1]{\endgroup#1\@@endlink}%
\providecommand \@sanitize@url [0]{\catcode `\\12\catcode `\$12\catcode
  `\&12\catcode `\#12\catcode `\^12\catcode `\_12\catcode `\%12\relax}%
\providecommand \@@startlink[1]{}%
\providecommand \@@endlink[0]{}%
\providecommand \url  [0]{\begingroup\@sanitize@url \@url }%
\providecommand \@url [1]{\endgroup\@href {#1}{\urlprefix }}%
\providecommand \urlprefix  [0]{URL }%
\providecommand \Eprint [0]{\href }%
\providecommand \doibase [0]{http://dx.doi.org/}%
\providecommand \selectlanguage [0]{\@gobble}%
\providecommand \bibinfo  [0]{\@secondoftwo}%
\providecommand \bibfield  [0]{\@secondoftwo}%
\providecommand \translation [1]{[#1]}%
\providecommand \BibitemOpen [0]{}%
\providecommand \bibitemStop [0]{}%
\providecommand \bibitemNoStop [0]{.\EOS\space}%
\providecommand \EOS [0]{\spacefactor3000\relax}%
\providecommand \BibitemShut  [1]{\csname bibitem#1\endcsname}%
\let\auto@bib@innerbib\@empty
\bibitem [{\citenamefont {{Abanov}}\ \emph {et~al.}(2003)\citenamefont
  {{Abanov}}, \citenamefont {{Chubukov}},\ and\ \citenamefont
  {{Schmalian}}}]{abanov:2003}%
  \BibitemOpen
  \bibfield  {author} {\bibinfo {author} {\bibfnamefont {A.}~\bibnamefont
  {{Abanov}}}, \bibinfo {author} {\bibfnamefont {A.~V.}\ \bibnamefont
  {{Chubukov}}}, \ and\ \bibinfo {author} {\bibfnamefont {J.}~\bibnamefont
  {{Schmalian}}},\ }\href {\doibase 10.1080/0001873021000057123} {\bibfield
  {journal} {\bibinfo  {journal} {Adv. Phys.}\ }\textbf {\bibinfo {volume}
  {52}},\ \bibinfo {pages} {119} (\bibinfo {year} {2003})}\BibitemShut
  {NoStop}%
\bibitem [{\citenamefont {{Abanov}}\ and\ \citenamefont
  {{Chubukov}}(2004)}]{abanov:2004}%
  \BibitemOpen
  \bibfield  {author} {\bibinfo {author} {\bibfnamefont {A.}~\bibnamefont
  {{Abanov}}}\ and\ \bibinfo {author} {\bibfnamefont {A.}~\bibnamefont
  {{Chubukov}}},\ }\href {\doibase 10.1103/PhysRevLett.93.255702} {\bibfield
  {journal} {\bibinfo  {journal} {\prl}\ }\textbf {\bibinfo {volume} {93}},\
  \bibinfo {eid} {255702} (\bibinfo {year} {2004})}\BibitemShut {NoStop}%
\bibitem [{\citenamefont {{Metlitski}}\ and\ \citenamefont
  {{Sachdev}}(2010{\natexlab{a}})}]{metlitski:2010b}%
  \BibitemOpen
  \bibfield  {author} {\bibinfo {author} {\bibfnamefont {M.~A.}\ \bibnamefont
  {{Metlitski}}}\ and\ \bibinfo {author} {\bibfnamefont {S.}~\bibnamefont
  {{Sachdev}}},\ }\href {\doibase 10.1103/PhysRevB.82.075127} {\bibfield
  {journal} {\bibinfo  {journal} {\prb}\ }\textbf {\bibinfo {volume} {82}},\
  \bibinfo {eid} {075127} (\bibinfo {year} {2010}{\natexlab{a}})}\BibitemShut
  {NoStop}%
\bibitem [{\citenamefont {{Metlitski}}\ and\ \citenamefont
  {{Sachdev}}(2010{\natexlab{b}})}]{metlitski:2010c}%
  \BibitemOpen
  \bibfield  {author} {\bibinfo {author} {\bibfnamefont {M.~A.}\ \bibnamefont
  {{Metlitski}}}\ and\ \bibinfo {author} {\bibfnamefont {S.}~\bibnamefont
  {{Sachdev}}},\ }\href {\doibase 10.1103/PhysRevB.82.075128} {\bibfield
  {journal} {\bibinfo  {journal} {\prb}\ }\textbf {\bibinfo {volume} {82}},\
  \bibinfo {eid} {075128} (\bibinfo {year} {2010}{\natexlab{b}})}\BibitemShut
  {NoStop}%
\bibitem [{\citenamefont {Holder}\ and\ \citenamefont
  {Metzner}(2015)}]{metzner:2015}%
  \BibitemOpen
  \bibfield  {author} {\bibinfo {author} {\bibfnamefont {T.}~\bibnamefont
  {Holder}}\ and\ \bibinfo {author} {\bibfnamefont {W.}~\bibnamefont
  {Metzner}},\ }\href {\doibase 10.1103/PhysRevB.92.245128} {\bibfield
  {journal} {\bibinfo  {journal} {Phys. Rev. B}\ }\textbf {\bibinfo {volume}
  {92}},\ \bibinfo {pages} {245128} (\bibinfo {year} {2015})}\BibitemShut
  {NoStop}%
\bibitem [{\citenamefont {{Basov}}\ \emph {et~al.}(2011)\citenamefont
  {{Basov}}, \citenamefont {{Averitt}}, \citenamefont {{van der Marel}},
  \citenamefont {{Dressel}},\ and\ \citenamefont {{Haule}}}]{basov:2011}%
  \BibitemOpen
  \bibfield  {author} {\bibinfo {author} {\bibfnamefont {D.~N.}\ \bibnamefont
  {{Basov}}}, \bibinfo {author} {\bibfnamefont {R.~D.}\ \bibnamefont
  {{Averitt}}}, \bibinfo {author} {\bibfnamefont {D.}~\bibnamefont {{van der
  Marel}}}, \bibinfo {author} {\bibfnamefont {M.}~\bibnamefont {{Dressel}}}, \
  and\ \bibinfo {author} {\bibfnamefont {K.}~\bibnamefont {{Haule}}},\ }\href
  {\doibase 10.1103/RevModPhys.83.471} {\bibfield  {journal} {\bibinfo
  {journal} {Rev. Mod. Phys.}\ }\textbf {\bibinfo {volume} {83}},\ \bibinfo
  {pages} {471} (\bibinfo {year} {2011})}\BibitemShut {NoStop}%
\bibitem [{\citenamefont {G\"otze}\ and\ \citenamefont
  {W\"olfle}(1972)}]{goetze:1972}%
  \BibitemOpen
  \bibfield  {author} {\bibinfo {author} {\bibfnamefont {W.}~\bibnamefont
  {G\"otze}}\ and\ \bibinfo {author} {\bibfnamefont {P.}~\bibnamefont
  {W\"olfle}},\ }\href {\doibase 10.1103/PhysRevB.6.1226} {\bibfield  {journal}
  {\bibinfo  {journal} {Phys. Rev. B}\ }\textbf {\bibinfo {volume} {6}},\
  \bibinfo {pages} {1226} (\bibinfo {year} {1972})}\BibitemShut {NoStop}%
\bibitem [{\citenamefont {Kim}\ \emph {et~al.}(1994)\citenamefont {Kim},
  \citenamefont {Furusaki}, \citenamefont {Wen},\ and\ \citenamefont
  {Lee}}]{kim:1994}%
  \BibitemOpen
  \bibfield  {author} {\bibinfo {author} {\bibfnamefont {Y.~B.}\ \bibnamefont
  {Kim}}, \bibinfo {author} {\bibfnamefont {A.}~\bibnamefont {Furusaki}},
  \bibinfo {author} {\bibfnamefont {X.-G.}\ \bibnamefont {Wen}}, \ and\
  \bibinfo {author} {\bibfnamefont {P.~A.}\ \bibnamefont {Lee}},\ }\href
  {\doibase 10.1103/PhysRevB.50.17917} {\bibfield  {journal} {\bibinfo
  {journal} {Phys. Rev. B}\ }\textbf {\bibinfo {volume} {50}},\ \bibinfo
  {pages} {17917} (\bibinfo {year} {1994})}\BibitemShut {NoStop}%
\bibitem [{\citenamefont {Eberlein}\ \emph {et~al.}(2016)\citenamefont
  {Eberlein}, \citenamefont {Mandal},\ and\ \citenamefont
  {Sachdev}}]{eberlein:2016}%
  \BibitemOpen
  \bibfield  {author} {\bibinfo {author} {\bibfnamefont {A.}~\bibnamefont
  {Eberlein}}, \bibinfo {author} {\bibfnamefont {I.}~\bibnamefont {Mandal}}, \
  and\ \bibinfo {author} {\bibfnamefont {S.}~\bibnamefont {Sachdev}},\ }\href
  {\doibase 10.1103/PhysRevB.94.045133} {\bibfield  {journal} {\bibinfo
  {journal} {Phys. Rev. B}\ }\textbf {\bibinfo {volume} {94}},\ \bibinfo
  {pages} {045133} (\bibinfo {year} {2016})}\BibitemShut {NoStop}%
\bibitem [{\citenamefont {Kim}(2016)}]{kim:private}%
  \BibitemOpen
  \bibfield  {author} {\bibinfo {author} {\bibfnamefont {Y.~B.}\ \bibnamefont
  {Kim}},\ }\href@noop {} {} (\bibinfo {year} {2016}),\ \bibinfo {note}
  {{private communication}}\BibitemShut {NoStop}%
\bibitem [{\citenamefont {Kim}\ \emph {et~al.}(1995)\citenamefont {Kim},
  \citenamefont {Lee},\ and\ \citenamefont {Wen}}]{kim:1995}%
  \BibitemOpen
  \bibfield  {author} {\bibinfo {author} {\bibfnamefont {Y.~B.}\ \bibnamefont
  {Kim}}, \bibinfo {author} {\bibfnamefont {P.~A.}\ \bibnamefont {Lee}}, \ and\
  \bibinfo {author} {\bibfnamefont {X.-G.}\ \bibnamefont {Wen}},\ }\href
  {\doibase 10.1103/PhysRevB.52.17275} {\bibfield  {journal} {\bibinfo
  {journal} {Phys. Rev. B}\ }\textbf {\bibinfo {volume} {52}},\ \bibinfo
  {pages} {17275} (\bibinfo {year} {1995})}\BibitemShut {NoStop}%
\bibitem [{\citenamefont {{Georges}}\ \emph {et~al.}(1996)\citenamefont
  {{Georges}}, \citenamefont {{Kotliar}}, \citenamefont {{Krauth}},\ and\
  \citenamefont {{Rozenberg}}}]{georges:1996}%
  \BibitemOpen
  \bibfield  {author} {\bibinfo {author} {\bibfnamefont {A.}~\bibnamefont
  {{Georges}}}, \bibinfo {author} {\bibfnamefont {G.}~\bibnamefont
  {{Kotliar}}}, \bibinfo {author} {\bibfnamefont {W.}~\bibnamefont {{Krauth}}},
  \ and\ \bibinfo {author} {\bibfnamefont {M.~J.}\ \bibnamefont
  {{Rozenberg}}},\ }\href {\doibase 10.1103/RevModPhys.68.13} {\bibfield
  {journal} {\bibinfo  {journal} {Rev. Mod. Phys.}\ }\textbf {\bibinfo {volume}
  {68}},\ \bibinfo {pages} {13} (\bibinfo {year} {1996})}\BibitemShut {NoStop}%
\bibitem [{\citenamefont {Hartnoll}\ \emph {et~al.}(2011)\citenamefont
  {Hartnoll}, \citenamefont {Hofman}, \citenamefont {Metlitski},\ and\
  \citenamefont {Sachdev}}]{hartnoll:2011}%
  \BibitemOpen
  \bibfield  {author} {\bibinfo {author} {\bibfnamefont {S.~A.}\ \bibnamefont
  {Hartnoll}}, \bibinfo {author} {\bibfnamefont {D.~M.}\ \bibnamefont
  {Hofman}}, \bibinfo {author} {\bibfnamefont {M.~A.}\ \bibnamefont
  {Metlitski}}, \ and\ \bibinfo {author} {\bibfnamefont {S.}~\bibnamefont
  {Sachdev}},\ }\href {\doibase 10.1103/PhysRevB.84.125115} {\bibfield
  {journal} {\bibinfo  {journal} {Phys. Rev. B}\ }\textbf {\bibinfo {volume}
  {84}},\ \bibinfo {pages} {125115} (\bibinfo {year} {2011})}\BibitemShut
  {NoStop}%
\bibitem [{\citenamefont {Chubukov}\ \emph {et~al.}(2014)\citenamefont
  {Chubukov}, \citenamefont {Maslov},\ and\ \citenamefont
  {Yudson}}]{chubukov:2014}%
  \BibitemOpen
  \bibfield  {author} {\bibinfo {author} {\bibfnamefont {A.~V.}\ \bibnamefont
  {Chubukov}}, \bibinfo {author} {\bibfnamefont {D.~L.}\ \bibnamefont
  {Maslov}}, \ and\ \bibinfo {author} {\bibfnamefont {V.~I.}\ \bibnamefont
  {Yudson}},\ }\href {\doibase 10.1103/PhysRevB.89.155126} {\bibfield
  {journal} {\bibinfo  {journal} {Phys. Rev. B}\ }\textbf {\bibinfo {volume}
  {89}},\ \bibinfo {pages} {155126} (\bibinfo {year} {2014})}\BibitemShut
  {NoStop}%
\bibitem [{\citenamefont {Rech}\ \emph {et~al.}(2006)\citenamefont {Rech},
  \citenamefont {P\'epin},\ and\ \citenamefont {Chubukov}}]{rech:2006}%
  \BibitemOpen
  \bibfield  {author} {\bibinfo {author} {\bibfnamefont {J.}~\bibnamefont
  {Rech}}, \bibinfo {author} {\bibfnamefont {C.}~\bibnamefont {P\'epin}}, \
  and\ \bibinfo {author} {\bibfnamefont {A.~V.}\ \bibnamefont {Chubukov}},\
  }\href {\doibase 10.1103/PhysRevB.74.195126} {\bibfield  {journal} {\bibinfo
  {journal} {Phys. Rev. B}\ }\textbf {\bibinfo {volume} {74}},\ \bibinfo
  {pages} {195126} (\bibinfo {year} {2006})}\BibitemShut {NoStop}%
\bibitem [{\citenamefont {{Chubukov}}(2005)}]{chubukov:2005d}%
  \BibitemOpen
  \bibfield  {author} {\bibinfo {author} {\bibfnamefont {A.~V.}\ \bibnamefont
  {{Chubukov}}},\ }\href {\doibase 10.1103/PhysRevB.71.245123} {\bibfield
  {journal} {\bibinfo  {journal} {\prb}\ }\textbf {\bibinfo {volume} {71}},\
  \bibinfo {eid} {245123} (\bibinfo {year} {2005})}\BibitemShut {NoStop}%
\bibitem [{\citenamefont {Maslov}\ \emph {et~al.}(2011)\citenamefont {Maslov},
  \citenamefont {Yudson},\ and\ \citenamefont {Chubukov}}]{maslov:2011}%
  \BibitemOpen
  \bibfield  {author} {\bibinfo {author} {\bibfnamefont {D.~L.}\ \bibnamefont
  {Maslov}}, \bibinfo {author} {\bibfnamefont {V.~I.}\ \bibnamefont {Yudson}},
  \ and\ \bibinfo {author} {\bibfnamefont {A.~V.}\ \bibnamefont {Chubukov}},\
  }\href {\doibase 10.1103/PhysRevLett.106.106403} {\bibfield  {journal}
  {\bibinfo  {journal} {\prl}\ }\textbf {\bibinfo {volume} {106}},\ \bibinfo
  {pages} {106403} (\bibinfo {year} {2011})}\BibitemShut {NoStop}%
\bibitem [{\citenamefont {Pal}\ \emph {et~al.}(2012)\citenamefont {Pal},
  \citenamefont {Yudson},\ and\ \citenamefont {Maslov}}]{pal:2012b}%
  \BibitemOpen
  \bibfield  {author} {\bibinfo {author} {\bibfnamefont {H.~K.}\ \bibnamefont
  {Pal}}, \bibinfo {author} {\bibfnamefont {V.~I.}\ \bibnamefont {Yudson}}, \
  and\ \bibinfo {author} {\bibfnamefont {D.~L.}\ \bibnamefont {Maslov}},\
  }\href {http://dx.doi.org/10.3952/lithjphys.52207} {\bibfield  {journal}
  {\bibinfo  {journal} {Lith. J. Phys.}\ }\textbf {\bibinfo {volume} {52}},\
  \bibinfo {pages} {{142}} (\bibinfo {year} {2012})}\BibitemShut {NoStop}%
\bibitem [{\citenamefont {Gurzhi}(1959)}]{gurzhi:1959}%
  \BibitemOpen
  \bibfield  {author} {\bibinfo {author} {\bibfnamefont {R.~N.}\ \bibnamefont
  {Gurzhi}},\ }\href@noop {} {\bibfield  {journal} {\bibinfo  {journal} {Sov.
  Phys.--JETP}\ }\textbf {\bibinfo {volume} {35}},\ \bibinfo {pages} {673}
  (\bibinfo {year} {1959})}\BibitemShut {NoStop}%
\bibitem [{\citenamefont {Maslov}\ and\ \citenamefont
  {Chubukov}(2017)}]{maslov:2017}%
  \BibitemOpen
  \bibfield  {author} {\bibinfo {author} {\bibfnamefont {D.~L.}\ \bibnamefont
  {Maslov}}\ and\ \bibinfo {author} {\bibfnamefont {A.~V.}\ \bibnamefont
  {Chubukov}},\ }\href {http://stacks.iop.org/0034-4885/80/i=2/a=026503}
  {\bibfield  {journal} {\bibinfo  {journal} {Rep. Prog. Phys.}\ }\textbf
  {\bibinfo {volume} {80}},\ \bibinfo {pages} {026503} (\bibinfo {year}
  {2017})}\BibitemShut {NoStop}%
\bibitem [{\citenamefont {Gurzhi}\ \emph {et~al.}(1982)\citenamefont {Gurzhi},
  \citenamefont {Kopeliovich},\ and\ \citenamefont {Rutkevich}}]{gurzhi:1982}%
  \BibitemOpen
  \bibfield  {author} {\bibinfo {author} {\bibfnamefont {R.}~\bibnamefont
  {Gurzhi}}, \bibinfo {author} {\bibfnamefont {A.}~\bibnamefont {Kopeliovich}},
  \ and\ \bibinfo {author} {\bibfnamefont {S.~B.}\ \bibnamefont {Rutkevich}},\
  }\href {http://jetp.ac.ru/cgi-bin/e/index/e/56/1/p159?a=list} {\bibfield
  {journal} {\bibinfo  {journal} {Sov. Phys.--JETP}\ }\textbf {\bibinfo
  {volume} {56}},\ \bibinfo {pages} {159} (\bibinfo {year} {1982})}\BibitemShut
  {NoStop}%
\bibitem [{\citenamefont {Rosch}\ and\ \citenamefont
  {Howell}(2005)}]{rosch:2005}%
  \BibitemOpen
  \bibfield  {author} {\bibinfo {author} {\bibfnamefont {A.}~\bibnamefont
  {Rosch}}\ and\ \bibinfo {author} {\bibfnamefont {P.~C.}\ \bibnamefont
  {Howell}},\ }\href {\doibase 10.1103/PhysRevB.72.104510} {\bibfield
  {journal} {\bibinfo  {journal} {Phys. Rev. B}\ }\textbf {\bibinfo {volume}
  {72}},\ \bibinfo {pages} {104510} (\bibinfo {year} {2005})}\BibitemShut
  {NoStop}%
\bibitem [{\citenamefont {Rosch}(2006)}]{rosch:2006}%
  \BibitemOpen
  \bibfield  {author} {\bibinfo {author} {\bibfnamefont {A.}~\bibnamefont
  {Rosch}},\ }\href {\doibase 10.1002/andp.200510203} {\bibfield  {journal}
  {\bibinfo  {journal} {Ann. Phys.}\ }\textbf {\bibinfo {volume} {15}},\
  \bibinfo {pages} {526} (\bibinfo {year} {2006})}\BibitemShut {NoStop}%
\bibitem [{\citenamefont {Briskot}\ \emph {et~al.}(2015)\citenamefont
  {Briskot}, \citenamefont {Sch\"utt}, \citenamefont {Gornyi}, \citenamefont
  {Titov}, \citenamefont {Narozhny},\ and\ \citenamefont
  {Mirlin}}]{briskot:2015}%
  \BibitemOpen
  \bibfield  {author} {\bibinfo {author} {\bibfnamefont {U.}~\bibnamefont
  {Briskot}}, \bibinfo {author} {\bibfnamefont {M.}~\bibnamefont {Sch\"utt}},
  \bibinfo {author} {\bibfnamefont {I.~V.}\ \bibnamefont {Gornyi}}, \bibinfo
  {author} {\bibfnamefont {M.}~\bibnamefont {Titov}}, \bibinfo {author}
  {\bibfnamefont {B.~N.}\ \bibnamefont {Narozhny}}, \ and\ \bibinfo {author}
  {\bibfnamefont {A.~D.}\ \bibnamefont {Mirlin}},\ }\href {\doibase
  10.1103/PhysRevB.92.115426} {\bibfield  {journal} {\bibinfo  {journal} {Phys.
  Rev. B}\ }\textbf {\bibinfo {volume} {92}},\ \bibinfo {pages} {115426}
  (\bibinfo {year} {2015})}\BibitemShut {NoStop}%
\bibitem [{\citenamefont {Holstein}(1964)}]{holstein:1964}%
  \BibitemOpen
  \bibfield  {author} {\bibinfo {author} {\bibfnamefont {T.}~\bibnamefont
  {Holstein}},\ }\href {\doibase
  http://dx.doi.org/10.1016/0003-4916(64)90008-9} {\bibfield  {journal}
  {\bibinfo  {journal} {Ann. Phys.}\ }\textbf {\bibinfo {volume} {29}},\
  \bibinfo {pages} {410 } (\bibinfo {year} {1964})}\BibitemShut {NoStop}%
\bibitem [{\citenamefont {Riseborough}(1983)}]{riseborough:1983}%
  \BibitemOpen
  \bibfield  {author} {\bibinfo {author} {\bibfnamefont {P.~S.}\ \bibnamefont
  {Riseborough}},\ }\href {\doibase 10.1103/PhysRevB.27.5775} {\bibfield
  {journal} {\bibinfo  {journal} {Phys. Rev. B}\ }\textbf {\bibinfo {volume}
  {27}},\ \bibinfo {pages} {5775} (\bibinfo {year} {1983})}\BibitemShut
  {NoStop}%
\bibitem [{\citenamefont {Yamada}\ and\ \citenamefont
  {Yosida}(1986)}]{yamada:1986}%
  \BibitemOpen
  \bibfield  {author} {\bibinfo {author} {\bibfnamefont {K.}~\bibnamefont
  {Yamada}}\ and\ \bibinfo {author} {\bibfnamefont {K.}~\bibnamefont
  {Yosida}},\ }\href {\doibase 10.1143/PTP.76.621} {\bibfield  {journal}
  {\bibinfo  {journal} {Prog. Theor. Phys.}\ }\textbf {\bibinfo {volume}
  {76}},\ \bibinfo {pages} {621} (\bibinfo {year} {1986})}\BibitemShut
  {NoStop}%
\bibitem [{\citenamefont {Gornyi}\ and\ \citenamefont
  {Mirlin}(2004)}]{gornyi:2004}%
  \BibitemOpen
  \bibfield  {author} {\bibinfo {author} {\bibfnamefont {I.~V.}\ \bibnamefont
  {Gornyi}}\ and\ \bibinfo {author} {\bibfnamefont {A.~D.}\ \bibnamefont
  {Mirlin}},\ }\href {\doibase 10.1103/PhysRevB.69.045313} {\bibfield
  {journal} {\bibinfo  {journal} {Phys. Rev. B}\ }\textbf {\bibinfo {volume}
  {69}},\ \bibinfo {pages} {045313} (\bibinfo {year} {2004})}\BibitemShut
  {NoStop}%
\bibitem [{\citenamefont {Berthod}\ \emph {et~al.}(2013)\citenamefont
  {Berthod}, \citenamefont {Mravlje}, \citenamefont {Deng}, \citenamefont
  {\ifmmode~\check{Z}\else \v{Z}\fi{}itko}, \citenamefont {van~der Marel},\
  and\ \citenamefont {Georges}}]{berthod:2013}%
  \BibitemOpen
  \bibfield  {author} {\bibinfo {author} {\bibfnamefont {C.}~\bibnamefont
  {Berthod}}, \bibinfo {author} {\bibfnamefont {J.}~\bibnamefont {Mravlje}},
  \bibinfo {author} {\bibfnamefont {X.}~\bibnamefont {Deng}}, \bibinfo {author}
  {\bibfnamefont {R.}~\bibnamefont {\ifmmode~\check{Z}\else \v{Z}\fi{}itko}},
  \bibinfo {author} {\bibfnamefont {D.}~\bibnamefont {van~der Marel}}, \ and\
  \bibinfo {author} {\bibfnamefont {A.}~\bibnamefont {Georges}},\ }\href
  {\doibase 10.1103/PhysRevB.87.115109} {\bibfield  {journal} {\bibinfo
  {journal} {Phys. Rev. B}\ }\textbf {\bibinfo {volume} {87}},\ \bibinfo
  {pages} {115109} (\bibinfo {year} {2013})}\BibitemShut {NoStop}%
\bibitem [{Note1()}]{Note1}%
  \BibitemOpen
  \bibinfo {note} {It can be shown that keeping terms of order $q$ in the
  integral equation for the current vertex gives corrections of order $\Omega
  /E_F$, which can be safely discarded.}\BibitemShut {Stop}%
\bibitem [{\citenamefont {Eliashberg}(1962)}]{eliashberg:1962}%
  \BibitemOpen
  \bibfield  {author} {\bibinfo {author} {\bibfnamefont {G.~M.}\ \bibnamefont
  {Eliashberg}},\ }\href {http://jetp.ac.ru/cgi-bin/e/index/e/14/4/p886?a=list}
  {\bibfield  {journal} {\bibinfo  {journal} {Sov. Phys.-JETP}\ }\textbf
  {\bibinfo {volume} {14}},\ \bibinfo {pages} {886} (\bibinfo {year}
  {1962})}\BibitemShut {NoStop}%
\bibitem [{\citenamefont {Ipatova}\ and\ \citenamefont
  {Eliashberg}(1962)}]{ipatova:1962}%
  \BibitemOpen
  \bibfield  {author} {\bibinfo {author} {\bibfnamefont {L.~P.}\ \bibnamefont
  {Ipatova}}\ and\ \bibinfo {author} {\bibfnamefont {G.~M.}\ \bibnamefont
  {Eliashberg}},\ }\href
  {http://jetp.ac.ru/cgi-bin/e/index/e/16/5/p1269?a=list} {\bibfield  {journal}
  {\bibinfo  {journal} {Sov. Phys. JETP}\ }\textbf {\bibinfo {volume} {16}},\
  \bibinfo {pages} {1269} (\bibinfo {year} {1962})}\BibitemShut {NoStop}%
\bibitem [{\citenamefont {Dzyaloshinskii}\ and\ \citenamefont
  {Larkin}(1972)}]{larkin:1972}%
  \BibitemOpen
  \bibfield  {author} {\bibinfo {author} {\bibfnamefont {I.~E.}\ \bibnamefont
  {Dzyaloshinskii}}\ and\ \bibinfo {author} {\bibfnamefont {A.~I.}\
  \bibnamefont {Larkin}},\ }\href
  {http://jetp.ac.ru/cgi-bin/e/index/e/34/2/p422?a=list} {\bibfield  {journal}
  {\bibinfo  {journal} {Sov. Phys. JETP}\ }\textbf {\bibinfo {volume} {34}},\
  \bibinfo {pages} {422} (\bibinfo {year} {1972})}\BibitemShut {NoStop}%
\bibitem [{\citenamefont {Shekhter}\ \emph {et~al.}(2005)\citenamefont
  {Shekhter}, \citenamefont {Khodas},\ and\ \citenamefont
  {Finkel'stein}}]{shekhter:2005}%
  \BibitemOpen
  \bibfield  {author} {\bibinfo {author} {\bibfnamefont {A.}~\bibnamefont
  {Shekhter}}, \bibinfo {author} {\bibfnamefont {M.}~\bibnamefont {Khodas}}, \
  and\ \bibinfo {author} {\bibfnamefont {A.~M.}\ \bibnamefont {Finkel'stein}},\
  }\href {\doibase 10.1103/PhysRevB.71.165329} {\bibfield  {journal} {\bibinfo
  {journal} {Phys. Rev. B}\ }\textbf {\bibinfo {volume} {71}},\ \bibinfo
  {pages} {165329} (\bibinfo {year} {2005})}\BibitemShut {NoStop}%
\bibitem [{\citenamefont {Chubukov}\ and\ \citenamefont
  {W\"olfle}(2014)}]{chubukov:2014b}%
  \BibitemOpen
  \bibfield  {author} {\bibinfo {author} {\bibfnamefont {A.~V.}\ \bibnamefont
  {Chubukov}}\ and\ \bibinfo {author} {\bibfnamefont {P.}~\bibnamefont
  {W\"olfle}},\ }\href {\doibase 10.1103/PhysRevB.89.045108} {\bibfield
  {journal} {\bibinfo  {journal} {Phys. Rev. B}\ }\textbf {\bibinfo {volume}
  {89}},\ \bibinfo {pages} {045108} (\bibinfo {year} {2014})}\BibitemShut
  {NoStop}%
\bibitem [{\citenamefont {Finkel'stein}(1983)}]{finkelshtein:1983}%
  \BibitemOpen
  \bibfield  {author} {\bibinfo {author} {\bibfnamefont {A.~M.}\ \bibnamefont
  {Finkel'stein}},\ }\href
  {http://jetp.ac.ru/cgi-bin/e/index/e/57/1/p97?a=list} {\bibfield  {journal}
  {\bibinfo  {journal} {Sov. Phys. JETP}\ }\textbf {\bibinfo {volume} {57}},\
  \bibinfo {pages} {97} (\bibinfo {year} {1983})}\BibitemShut {NoStop}%
\bibitem [{\citenamefont {Finkel'stein}(2010)}]{finkelshtein:2010}%
  \BibitemOpen
  \bibfield  {author} {\bibinfo {author} {\bibfnamefont {A.~M.}\ \bibnamefont
  {Finkel'stein}},\ }\href {\doibase 10.1142/S0217979210064642} {\bibfield
  {journal} {\bibinfo  {journal} {Int. J. Mod. Phys. B}\ }\textbf {\bibinfo
  {volume} {24}},\ \bibinfo {pages} {1855} (\bibinfo {year}
  {2010})}\BibitemShut {NoStop}%
\bibitem [{\citenamefont {{A. A. Abrikosov, L. P. Gorkov, and I. E.
  Dzyaloshinski}}(1963)}]{agd:1963}%
  \BibitemOpen
  \bibfield  {author} {\bibinfo {author} {\bibnamefont {{A. A. Abrikosov, L. P.
  Gorkov, and I. E. Dzyaloshinski}}},\ }\href@noop {} {\emph {\bibinfo {title}
  {Methods of Quantum Field Theory in Statistical Physics}}}\ (\bibinfo
  {publisher} {Dover, New York},\ \bibinfo {year} {1963})\BibitemShut {NoStop}%
\bibitem [{\citenamefont {Chubukov}(2005)}]{chubukov:2005c}%
  \BibitemOpen
  \bibfield  {author} {\bibinfo {author} {\bibfnamefont {A.~V.}\ \bibnamefont
  {Chubukov}},\ }\href {\doibase 10.1103/PhysRevB.72.085113} {\bibfield
  {journal} {\bibinfo  {journal} {Phys. Rev. B}\ }\textbf {\bibinfo {volume}
  {72}},\ \bibinfo {pages} {085113} (\bibinfo {year} {2005})}\BibitemShut
  {NoStop}%
\bibitem [{Note2()}]{Note2}%
  \BibitemOpen
  \bibinfo {note} {A frequency-independent conductivity in $D=2$ would be
  consistent with the hypothesis of hyperscaling,\cite {patel:2015} according
  to which $\sigma \propto \Omega ^{\protect \frac {D-2}{z}}$, where $z$ is the
  dynamical critical exponent. This agreement may, however, be accidental as it
  does not hold, within the same reasoning, beyond $D=2$.}\BibitemShut {Stop}%
\bibitem [{\citenamefont {Maslov}(2004)}]{maslov:2004}%
  \BibitemOpen
  \bibfield  {author} {\bibinfo {author} {\bibfnamefont {D.~L.}\ \bibnamefont
  {Maslov}},\ }in\ \href {https://arxiv.org/abs/cond-mat/0506035} {\emph
  {\bibinfo {booktitle} {Nanophysics: Coherence and Transport}}},\ \bibinfo
  {editor} {edited by\ \bibinfo {editor} {\bibfnamefont {G.~M.}\ \bibnamefont
  {H.~Bouchiat}, \bibfnamefont {Y.~Gefen}}\ and\ \bibinfo {editor}
  {\bibfnamefont {J.}~\bibnamefont {Dalibard}}}\ (\bibinfo  {publisher}
  {Elsevier, Amsterdam},\ \bibinfo {year} {2004})\ pp.\ \bibinfo {pages}
  {1--108}\BibitemShut {NoStop}%
\bibitem [{\citenamefont {Varma}\ \emph {et~al.}(1989)\citenamefont {Varma},
  \citenamefont {Littlewood}, \citenamefont {Schmitt-Rink}, \citenamefont
  {Abrahams},\ and\ \citenamefont {Ruckenstein}}]{varma:1989}%
  \BibitemOpen
  \bibfield  {author} {\bibinfo {author} {\bibfnamefont {C.~M.}\ \bibnamefont
  {Varma}}, \bibinfo {author} {\bibfnamefont {P.~B.}\ \bibnamefont
  {Littlewood}}, \bibinfo {author} {\bibfnamefont {S.}~\bibnamefont
  {Schmitt-Rink}}, \bibinfo {author} {\bibfnamefont {E.}~\bibnamefont
  {Abrahams}}, \ and\ \bibinfo {author} {\bibfnamefont {A.~E.}\ \bibnamefont
  {Ruckenstein}},\ }\href {\doibase 10.1103/PhysRevLett.63.1996} {\bibfield
  {journal} {\bibinfo  {journal} {Phys. Rev. Lett.}\ }\textbf {\bibinfo
  {volume} {63}},\ \bibinfo {pages} {1996} (\bibinfo {year}
  {1989})}\BibitemShut {NoStop}%
\bibitem [{\citenamefont {{Varma}}\ \emph {et~al.}(2002)\citenamefont
  {{Varma}}, \citenamefont {{Nussinov}},\ and\ \citenamefont {{van
  Saarloos}}}]{varma:2002}%
  \BibitemOpen
  \bibfield  {author} {\bibinfo {author} {\bibfnamefont {C.~M.}\ \bibnamefont
  {{Varma}}}, \bibinfo {author} {\bibfnamefont {Z.}~\bibnamefont {{Nussinov}}},
  \ and\ \bibinfo {author} {\bibfnamefont {W.}~\bibnamefont {{van Saarloos}}},\
  }\href {\doibase 10.1016/S0370-1573(01)00060-6} {\bibfield  {journal}
  {\bibinfo  {journal} {Phys. Rep.}\ }\textbf {\bibinfo {volume} {361}},\
  \bibinfo {pages} {267} (\bibinfo {year} {2002})}\BibitemShut {NoStop}%
\bibitem [{\citenamefont {{Basov}}\ \emph {et~al.}(1996)\citenamefont
  {{Basov}}, \citenamefont {{Liang}}, \citenamefont {{Dabrowski}},
  \citenamefont {{Bonn}}, \citenamefont {{Hardy}},\ and\ \citenamefont
  {{Timusk}}}]{basov:1996}%
  \BibitemOpen
  \bibfield  {author} {\bibinfo {author} {\bibfnamefont {D.~N.}\ \bibnamefont
  {{Basov}}}, \bibinfo {author} {\bibfnamefont {R.}~\bibnamefont {{Liang}}},
  \bibinfo {author} {\bibfnamefont {B.}~\bibnamefont {{Dabrowski}}}, \bibinfo
  {author} {\bibfnamefont {D.~A.}\ \bibnamefont {{Bonn}}}, \bibinfo {author}
  {\bibfnamefont {W.~N.}\ \bibnamefont {{Hardy}}}, \ and\ \bibinfo {author}
  {\bibfnamefont {T.}~\bibnamefont {{Timusk}}},\ }\href {\doibase
  10.1103/PhysRevLett.77.4090} {\bibfield  {journal} {\bibinfo  {journal}
  {\prl}\ }\textbf {\bibinfo {volume} {77}},\ \bibinfo {pages} {4090} (\bibinfo
  {year} {1996})}\BibitemShut {NoStop}%
\bibitem [{\citenamefont {{Puchkov}}\ \emph {et~al.}(1996)\citenamefont
  {{Puchkov}}, \citenamefont {{Basov}},\ and\ \citenamefont
  {{Timusk}}}]{puchkov:1996}%
  \BibitemOpen
  \bibfield  {author} {\bibinfo {author} {\bibfnamefont {A.~V.}\ \bibnamefont
  {{Puchkov}}}, \bibinfo {author} {\bibfnamefont {D.~N.}\ \bibnamefont
  {{Basov}}}, \ and\ \bibinfo {author} {\bibfnamefont {T.}~\bibnamefont
  {{Timusk}}},\ }\href {\doibase 10.1088/0953-8984/8/48/023} {\bibfield
  {journal} {\bibinfo  {journal} {J. Phys.: Condens. Matter}\ }\textbf
  {\bibinfo {volume} {8}},\ \bibinfo {pages} {10049} (\bibinfo {year}
  {1996})}\BibitemShut {NoStop}%
\bibitem [{\citenamefont {{Norman}}\ and\ \citenamefont
  {{Chubukov}}(2006)}]{norman:2006}%
  \BibitemOpen
  \bibfield  {author} {\bibinfo {author} {\bibfnamefont {M.~R.}\ \bibnamefont
  {{Norman}}}\ and\ \bibinfo {author} {\bibfnamefont {A.~V.}\ \bibnamefont
  {{Chubukov}}},\ }\href {\doibase 10.1103/PhysRevB.73.140501} {\bibfield
  {journal} {\bibinfo  {journal} {\prb}\ }\textbf {\bibinfo {volume} {73}},\
  \bibinfo {eid} {140501} (\bibinfo {year} {2006})}\BibitemShut {NoStop}%
\bibitem [{\citenamefont {Langer}(1960)}]{langer:1960}%
  \BibitemOpen
  \bibfield  {author} {\bibinfo {author} {\bibfnamefont {J.~S.}\ \bibnamefont
  {Langer}},\ }\href {\doibase 10.1103/PhysRev.120.714} {\bibfield  {journal}
  {\bibinfo  {journal} {Phys. Rev.}\ }\textbf {\bibinfo {volume} {120}},\
  \bibinfo {pages} {714} (\bibinfo {year} {1960})}\BibitemShut {NoStop}%
\bibitem [{\citenamefont {Langreth}\ and\ \citenamefont
  {Kadanoff}(1964)}]{langreth:1964}%
  \BibitemOpen
  \bibfield  {author} {\bibinfo {author} {\bibfnamefont {D.~C.}\ \bibnamefont
  {Langreth}}\ and\ \bibinfo {author} {\bibfnamefont {L.~P.}\ \bibnamefont
  {Kadanoff}},\ }\href {\doibase 10.1103/PhysRev.133.A1070} {\bibfield
  {journal} {\bibinfo  {journal} {Phys. Rev.}\ }\textbf {\bibinfo {volume}
  {133}},\ \bibinfo {pages} {A1070} (\bibinfo {year} {1964})}\BibitemShut
  {NoStop}%
\bibitem [{\citenamefont {{P. Nozi{\`e}res and D.
  Pines}}(1966)}]{nozieres:1966}%
  \BibitemOpen
  \bibfield  {author} {\bibinfo {author} {\bibnamefont {{P. Nozi{\`e}res and D.
  Pines}}},\ }\href@noop {} {\emph {\bibinfo {title} {{The Theory of Quantum
  Liquids}}}},\ Vol.~\bibinfo {volume} {1}\ (\bibinfo  {publisher} {New York:
  Benjamin},\ \bibinfo {year} {1966})\BibitemShut {NoStop}%
\bibitem [{\citenamefont {Lifshitz}\ and\ \citenamefont
  {Pitaevskii}(1981)}]{physkin}%
  \BibitemOpen
  \bibfield  {author} {\bibinfo {author} {\bibfnamefont {E.~M.}\ \bibnamefont
  {Lifshitz}}\ and\ \bibinfo {author} {\bibfnamefont {L.~P.}\ \bibnamefont
  {Pitaevskii}},\ }\href@noop {} {\emph {\bibinfo {title} {Physical
  Kinetics}}}\ (\bibinfo  {publisher} {Butterworth-Heinemann, Burlington},\
  \bibinfo {year} {1981})\BibitemShut {NoStop}%
\bibitem [{Note3()}]{Note3}%
  \BibitemOpen
  \bibinfo {note} {In the original derivation,\cite {nozieres:1966} $\gamma
  _{\protect \text {tr}}$ is due to impurity scattering but we adopt the same
  form, assuming momentarily that it holds also for other types of
  scattering.}\BibitemShut {Stop}%
\bibitem [{\citenamefont {Engelsberg}\ and\ \citenamefont
  {Schrieffer}(1963)}]{engelsberg:1963}%
  \BibitemOpen
  \bibfield  {author} {\bibinfo {author} {\bibfnamefont {S.}~\bibnamefont
  {Engelsberg}}\ and\ \bibinfo {author} {\bibfnamefont {J.~R.}\ \bibnamefont
  {Schrieffer}},\ }\href {\doibase 10.1103/PhysRev.131.993} {\bibfield
  {journal} {\bibinfo  {journal} {Phys. Rev.}\ }\textbf {\bibinfo {volume}
  {131}},\ \bibinfo {pages} {993} (\bibinfo {year} {1963})}\BibitemShut
  {NoStop}%
\bibitem [{\citenamefont {{Lee}}(2009)}]{lee:2009}%
  \BibitemOpen
  \bibfield  {author} {\bibinfo {author} {\bibfnamefont {S.-S.}\ \bibnamefont
  {{Lee}}},\ }\href {\doibase 10.1103/PhysRevB.80.165102} {\bibfield  {journal}
  {\bibinfo  {journal} {\prb}\ }\textbf {\bibinfo {volume} {80}},\ \bibinfo
  {eid} {165102} (\bibinfo {year} {2009})}\BibitemShut {NoStop}%
\bibitem [{\citenamefont {Patel}\ \emph {et~al.}(2015)\citenamefont {Patel},
  \citenamefont {Strack},\ and\ \citenamefont {Sachdev}}]{patel:2015}%
  \BibitemOpen
  \bibfield  {author} {\bibinfo {author} {\bibfnamefont {A.~A.}\ \bibnamefont
  {Patel}}, \bibinfo {author} {\bibfnamefont {P.}~\bibnamefont {Strack}}, \
  and\ \bibinfo {author} {\bibfnamefont {S.}~\bibnamefont {Sachdev}},\ }\href
  {\doibase 10.1103/PhysRevB.92.165105} {\bibfield  {journal} {\bibinfo
  {journal} {Phys. Rev. B}\ }\textbf {\bibinfo {volume} {92}},\ \bibinfo
  {pages} {165105} (\bibinfo {year} {2015})}\BibitemShut {NoStop}%
\end{thebibliography}%
\end{document}